\documentclass[preprint,showpacs,preprintnumbers,amsmath,amssymb]{revtex4}
\usepackage{graphicx}
\usepackage{dcolumn}
\usepackage{bm}
\usepackage{amssymb}
\usepackage{amsmath}
\usepackage{latexsym}

\begin{document}

\title{Open
quantum system in external magnetic field  within  non-Markovian quantum Langevin approach}
\author{
I.B. Abdurakhmanov$^{1}$, Z. Kanokov$^{2,3,4}$
G.G. Adamian$^{2}$, and N.V. Antonenko$^{2}$,
}
\affiliation{
$^{1}$Curtin Institute for Computation, Department of Physics, Astronomy and Medical Radiation Sciences, Curtin University,  Perth, WA 6845, Australia\\
$^{2}$Joint Institute for
Nuclear Research, 141980 Dubna, Russia\\
$^{3}$National University, 700174 Tashkent,
Uzbekistan\\
$^{4}$Institute of Nuclear Physics,
702132 Tashkent, Uzbekistan
}

\date{\today}

\begin{abstract}
The non-Markovian
dynamics of a  charged   particle
linearly coupled to a neutral bosonic heat bath is investigated in an external uniform magnetic field.
The analytical expressions for the time-dependent and asymptotic
friction and diffusion coefficients, cyclotron frequencies,
variances of the  coordinate and momentum, and  orbital magnetic moments are derived.
The role of magnetic field in the dissipation and
diffusion processes is illustrated by several examples in the low- and
high-temperature  regimes. The localization phenomenon for a charged particle is observed.
The orbital diamagnetism of quantum system in a dissipative environment is studied.
The quantization conditions are found for the angular momentum.
%

\end{abstract}

\maketitle
\section{ Introduction}
The problem of description of a
two-dimensional quantum system under the influence of  external
magnetic and electric fields and energy exchange with its environment
is of great interest in
 atomic, nuclear, and plasma physics, astrophysics, condensed matter
physics, quantum optics, and
quantum information and measurement theories \cite{knigaMenskogo,Fried,Chuvil,PS,Naz,Naz2,Langer,Ginz}. The intensive
investigations deal with the impact of the external magnetic field
on such systems as quantum dots, quantum wires, and
two-dimensional electronic systems \cite{PS}.  The characteristics
of plasma in the homogeneous external field has also importance in the
physics of gas discharge \cite{Ginz}.

The external fields modify the distribution
of   internal energy which in turn modifies or alters the electronic properties  such as
the carrier concentration, direction, and mobility.   Modeling
electric current implies the determination of the time-dependence of the number
of electrons with  given momentum at a certain
location. The equations of motion  can be obtained by using the quantum
Langevin approach or density matrix formalism
which is widely applied to find the effects of fluctuations and
dissipation in macroscopic systems
\cite{Kampen1,Kampen2,Kampen3,Kampen4,Kampen5,Kampen6,LEG,Dodonov,DM,Hu1,Hu2,Ma,Katia,Isar,Ford,Ford1,In,M110,Kanokov,Kanokov2,PA,Lac11,Lac13,Lac14}.
Using the phenomenological Markovian Fokker-Planck
equation for the Wigner probability function, the problem of
quantum description of the damped isotropic two-dimensional harmonic
oscillator in an uniform magnetic field has been studied in
Ref.~\cite{Dodonov} in the case of arbitrary relations between the
proper oscillator frequency,   damping coefficients and
temperature. The relations between the phenomenological diffusion
coefficients ensuring the positivity of the reduced density
matrix at each moment of time have been obtained in Ref.~\cite{Dodonov}.
By including
the magnetic field in the quantum non-Markovian Langevin equation,
the effects of dissipation and magnetic field on
localization of a charged particle moving in a confined potential  have been  investigated in Refs.~\cite{Ford,Ford1}.
As found, the weak
dissipation delocalizes the oscillation of  a charged particle
when the magnetic field is stronger than a certain critical value \cite{Ford1}.
For a charged particle moving in a two-dimensional harmonic oscillator and an uniform
  magnetic field,  the time-dependent
friction and diffusion coefficients have been  analytically
derived and numerically
studied within the non-Markovian quantum Langevin formalism in Ref. \cite{Kanokov2,PA}.
A  charged particle moving in a static external magnetic field (without a confined potential)
and linearly coupled to a heat bath  has been only treated in Refs.~\cite{Ma,In}, where
a fully dynamical calculation of the orbital diamagnetism has been presented.
In Ref.  \cite{Ma}
The non-Markovian and Markovian Langevin formalism have been used
  in Refs. \cite{In} and  \cite{Ma}, respectively.
In all cases \cite{Ma,Ford,Ford1,In,Kanokov2},
the magnetic field affects  neither   the memory function nor the random
force appearing in the quantum Langevin equation.

The aim of the present work is
to derive the analytical by the transport coefficients  for a unconfined charged particle
in an uniform magnetic field and dissipative environment and to
study the influence of the magnetic field on these
coefficients,  fluctuations,
cyclotron frequencies, and orbital magnetic moment (orbital diamagnetism).
The paper is organized as follows.
In Sec.~II, we give the Hamiltonian of
the system and solve the
quantum non-Markovian two-dimensional
Langevin equations for a charged particle
moving in the plane normal to the applied field.
The transport coefficients are obtained
by considering the first and second moments of the stochastic
dissipative equations.
The discussions and illustrative numerical
results are presented in Sec.~III.
A summary is given in Sec. IV.


\section{Linear coupling in coordinate with heat bath}
\subsection{Derivation of quantum Langevin equations}
Let us consider a two-dimensional  motion of  a charged quantum particle in the presence
of heat bath and
 external constant
 magnetic field ${\bf B}=(0,0,B)$.
 The total
Hamiltonian of this system is \cite{Ford,PA}
\begin{eqnarray}
H=H_c+H_b+H_{cb}.
\label{equ1}
\end{eqnarray}
The Hamiltonian $H_c$   describes the    charged quantum particle  with
effective mass tensor and charge $e=|e|$  in
magnetic field:
\begin{eqnarray}
H_c
= \frac{1}{2 m_x}[p_x-e A_x(x,y)]^2+\frac{1}{2 m_y}[p_y-e
A_y(x,y)]^2
= \frac{\pi_x^2}{2m_x}+\frac{\pi_y^2}{2m_y}.
\label{equ2}
\end{eqnarray}
Here,  $m_x$ and $m_y$ are the components of the effective mass
tensor, ${\bf R}=(x,y,0)$ and
${\bf p}=(p_x,p_y,0)$ are the coordinate and canonically conjugated momentum,
respectively, ${\bf A}=(-\frac{1}{2}y B,\frac{1}{2}x B,0)$ is
 the vector potential of the magnetic field.
For simplicity, in Eq.~(\ref{equ2})
we introduce  the notations
$$\pi_x=p_x+\frac{1}{2}m_x\omega_{cx}y,\hspace{.3in}
\pi_y=p_y-\frac{1}{2}m_y\omega_{cy}x$$ with  frequencies
 $\omega_{cx}=\frac{e B}{m_x }$ and $\omega_{cy}=\frac{e
B}{m_y }$.
The cyclotron frequency is $\omega_c=\sqrt{\omega_{cx} \omega_{cy}}=\frac{e
B}{\sqrt{m_x m_y}}$.

The second term in Eq.~(\ref{equ1}) represents the Hamiltonian of the phonon (bosonic)
heat bath,
\begin{eqnarray}
H_b=\sum_{\nu}^{}\hbar\omega_\nu b_\nu^+b_\nu,
\label{equ3}
\end{eqnarray}
where $b_\nu^+$ and $b_\nu$ are the phonon creation and
annihilation operators of the heat bath. The coupling between the
heat bath and   charged particle is described by
\begin{eqnarray}
H_{cb}&=&\sum_{\nu}^{}(\alpha_{\nu}x + \beta_{\nu}y)(b_\nu^+ +
b_\nu)+\sum_{\nu}^{}\frac{1}{\hbar\omega_\nu}(\alpha_{\nu}x + \beta_{\nu}y)^2,
\label{equ4}
\end{eqnarray}
where $\alpha_{\nu}$ and $\beta_{\nu}$ are the  real coupling
constants. Equation (\ref{equ4}) is already used in literature \cite{Ma,Ford,In,Kanokov2,PA}.
The first term of $H_{cb}$ in Eq. (\ref{equ4}) corresponds to the  energy  exchange   between
the charged particle and heat bath. We introduce the
counter-term (second term) in $H_{cb}$ in order to
compensate the coupling-induced  potential.
 In general case, $\alpha_{\nu}$ and $\beta_{\nu}$  depend  on the strength of
magnetic field and an impact of the magnetic field $\bf B$ is entered into the dissipative kernels
and random forces.

The equations of motion are
\begin{eqnarray}
\dot{x}(t)&=&\frac{i}{\hbar}[H,x]=\frac{\pi_x(t)}{m_x},\hspace{.3in}
\dot {y}(t)=\frac{i}{\hbar}[H,y]=\frac{\pi_y(t)}{m_y},\nonumber\\
\dot\pi_{x}(t)&=&\frac{i}{\hbar}[H,\pi_x]=\omega_{cy}\pi_{y}(t)
-\sum_{\nu}^{}\alpha_{\nu}(b_\nu^+(t) + b_\nu(t))-
2\sum_{\nu}^{}\frac{\alpha_{\nu}(\alpha_{\nu}x(t) + \beta_{\nu}y(t))}{\hbar\omega_{\nu}},\nonumber\\
\dot \pi_{y}(t)&=&\frac{i}{\hbar}[H,\pi_y]=-\omega_{cx}\pi_{x}(t)
-\sum_{\nu}^{}\beta_{\nu}(b_\nu^+(t) + b_\nu(t))-
2\sum_{\nu}^{}\frac{\beta_{\nu}
(\alpha_{\nu}x(t) + \beta_{\nu}y(t))}{\hbar\omega_{\nu}},
\label{equ5}
\end{eqnarray}
and
\begin{eqnarray}
\dot b_\nu^+(t)&=&\frac{i}{\hbar}[H,b_\nu^+]=
i\omega_\nu b_\nu^+(t) + \frac{i}{\hbar}(\alpha_{\nu}x(t) + \beta_{\nu}y(t)), \nonumber\\
\dot b_\nu(t)&=&\frac{i}{\hbar}[H,b_\nu]= -i\omega_\nu b_\nu(t)
- \frac{i}{\hbar}(\alpha_{\nu}x(t) + \beta_{\nu}y(t)).
\label{equ6}
\end{eqnarray}
The solution of Eqs.~(\ref{equ6}) are
\begin{eqnarray}
b_\nu^+(t)+b_\nu (t)&=&f^{+}_\nu (t)+f_\nu (t) -
\frac{2(\alpha_{\nu}x(t) + \beta_{\nu}y(t))}{\hbar\omega_\nu}  + \frac{2}{\hbar\omega_\nu}\int\limits_{0}^{t}d\tau
(\alpha_{\nu}{\dot x}(\tau) + \beta_{\nu}{\dot y}(\tau))\cos(\omega_\nu[t-\tau]),\nonumber\\
b_\nu^+(t)-b_\nu (t)&=&f^{+}_\nu (t)-f_\nu (t) +\frac{2i}{\hbar\omega_\nu}\int\limits_{0}^{t}d\tau
(\alpha_{\nu}{\dot x}(\tau) + \beta_{\nu}{\dot y}(\tau))\sin(\omega_\nu[t-\tau]),
\label{equ7}
\end{eqnarray}
where
\begin{eqnarray}
f_\nu(t)=\left[b_\nu(0)+\frac{\alpha_{\nu}x(0) + \beta_{\nu}y(0)}{\hbar\omega_\nu}\right]e^{-i\omega_\nu t}. \nonumber
\end{eqnarray}
Substituting (\ref{equ7})
into (\ref{equ5}) and  eliminating the bath variables from
the equations of motion for the charged particle, we obtain the
set of nonlinear integro-differential stochastic dissipative
equations
\begin{eqnarray}
\dot {x}(t)&=&\frac{\pi_x(t)}{m_x}, \hspace{.3in}
\dot {y}(t)=\frac{\pi_y(t)}{m_y}, \nonumber\\
\dot \pi_{x}(t)&=&\omega_{cy}\pi_{y}(t)
- \int\limits_{0}^{t}d\tau K_{xx}(t,\tau)\dot{x}(\tau)
 -\int\limits_{0}^{t}d\tau K_{xy}(t,\tau)\dot{x}(\tau)+ F_{x}(t),\nonumber\\
\dot \pi_{y}(t)&=&-\omega_{cx}\pi_{x}(t)  -
\int\limits_{0}^{t}d\tau K_{yy}(t,\tau)\dot{y}(\tau)
 -\int\limits_{0}^{t}d\tau K_{yx}(t,\tau)\dot{y}(\tau)+ F_{y}(t).
\label{equ8}
\end{eqnarray}
The dissipative kernels and random forces in (\ref{equ8}) are
\begin{eqnarray}
K_{xx}(t,\tau)&=& 2\sum_{\nu}^{}\frac{\alpha_{\nu}^2}{\hbar\omega_\nu}\cos(\omega_\nu [t-\tau]), \nonumber\\
K_{xy}(t,\tau)&=&K_{yx}(t,\tau)= 2\sum_{\nu}^{}\frac{\alpha_{\nu}\beta_{\nu}}{\hbar\omega_\nu}\cos(\omega_\nu [t-\tau]), \nonumber\\
K_{yy}(t,\tau)&=&\sum_{\nu}^{}\frac{\beta_{\nu}^2}{\hbar\omega_\nu}\cos(\omega_\nu[t-\tau])
%
\label{equ9}
\end{eqnarray}
and
\begin{eqnarray}
F_{x}(t)&=&\sum_{\nu}{}F_{x}^{\nu}(t)=-\sum_{\nu}{}\alpha_{\nu}[f_{\nu}^{+}(t)+f_{\nu}(t)],\nonumber\\
F_{y}(t)&=&\sum_{\nu}{}F_{y}^{\nu}(t)=-\sum_{\nu}{}\beta_{\nu}[f_{\nu}^{+}(t)+f_{\nu}(t)],
\label{equ10}
\end{eqnarray}
respectively.
Following the standard procedure of statistical mechanics,
we identify the  operators $F_{x}^{\nu}$ and $F_{y}^{\nu}$ as fluctuations because of
uncertainty of the initial conditions for the bath operators. To
specify the statistical properties of the fluctuations, we
consider an ensemble of initial states in which the fluctuations
have the Gaussian distribution with zero average value
\begin{eqnarray}
\ll F^\nu_x (t)\gg = \ll F^\nu_y (t)\gg = 0.
\label{equ11}
\end{eqnarray}
Here, the symbol $\ll...\gg$ denotes the  average over the bath with the
Bose-Einstein statistics
\begin{eqnarray}
\ll f_{\nu}^{+}(t)f_{\nu'}^{+}(t')\gg &=&\ll f_{\nu}(t)f_{\nu'}(t')\gg=0, \nonumber\\
\ll f_{\nu}^{+}(t)f_{\nu'}(t')\gg &=&\delta_{\nu,\nu'}n_{\nu}e^{i\omega_\nu [t-t']},\nonumber\\
\ll f_{\nu}(t)f_{\nu'}^{+}(t')\gg
&=&\delta_{\nu,\nu'}(n_{\nu}+1)e^{-i\omega_\nu [t-t']},
\label{equ12}
\end{eqnarray}
where the occupation numbers $n_\nu=[\exp(\hbar\omega_\nu/T)-1]^{-1}$ for phonons depend  on temperature
$T$ given in the energy units.

Using the properties   (\ref{equ11}) and (\ref{equ12}) of random forces,
we get the following
symmetrized correlation functions $\varphi_{kk'}^{\nu}(t,t')=\ll
F^{\nu}_k(t)F^{\nu}_{k'}(t')+F^{\nu}_{k'}(t')F^{\nu}_k(t)\gg$
($k, k'=x, y$):
\begin{eqnarray}
\varphi_{xx}^\nu(t,t')&=&2[2n_{\nu}+1]\alpha_{\nu}^2\cos(\omega_\nu [t-t']),\nonumber\\
\varphi_{yy}^\nu(t,t')&=&\varphi_{xx}^\nu(t,t')|_{{x\to y}},\nonumber\\
\varphi_{xy}^\nu(t,t')&=&2[2n_{\nu}+1]\alpha_{\nu}\beta_{\nu}\cos(\omega_\nu [t-t']),\nonumber\\
\varphi_{yx}^\nu(t,t')&=&\varphi_{xy}^\nu(t,t')|_{{x\to y}}.
\label{equ13}
\end{eqnarray}
The quantum fluctuation-dissipation relations read
\begin{eqnarray}
\sum_{\nu}^{}\varphi_{xx}^\nu(t,t')\frac{\tanh[\frac{\hbar\omega_\nu}{2T}]}{\hbar\omega_\nu}=K_{xx}(t,t'), \nonumber\\
\sum_{\nu}^{}\varphi_{yy}^\nu(t,t')\frac{\tanh[\frac{\hbar\omega_\nu}{2T}]}{\hbar\omega_\nu}=K_{yy}(t,t'),\nonumber\\
\sum_{\nu}^{}\varphi_{xy}^\nu(t,t')\frac{\tanh[\frac{\hbar\omega_\nu}{2T}]}{\hbar\omega_\nu}=K_{xy}(t,t'),\nonumber\\
\sum_{\nu}^{}\varphi_{yx}^\nu(t,t')\frac{\tanh[\frac{\hbar\omega_\nu}{2T}]}{\hbar\omega_\nu}=K_{yx}(t,t').
\label{equ14}
\end{eqnarray}
The validity of the fluctuation-dissipation relations means that we have properly identified the
dissipative terms in the non-Markovian dynamical equations of motion.
The quantum fluctuation-dissipation relations differ from the
classical ones 
\begin{eqnarray}
\sum_{\nu}^{}\varphi_{xx}^\nu(t,t')=2TK_{xx}(t,t'), \nonumber\\
\sum_{\nu}^{}\varphi_{yy}^\nu(t,t')=2TK_{yy}(t,t'),\nonumber\\
\sum_{\nu}^{}\varphi_{xy}^\nu(t,t')=2TK_{xy}(t,t'),\nonumber\\
\sum_{\nu}^{}\varphi_{yx}^\nu(t,t')=2TK_{yx}(t,t').
\label{equ14c}
\end{eqnarray}
and are reduced to them in the limit of high temperature.

\subsection{Solution of Non-Markovian Langevin equations}

In order to solve the equations of motion (\ref{equ8}) for the
variables of the charged particle, we applied the Laplace transformation which significantly
simplifies  the problem \cite{Kanokov,Kanokov2}.
The explicit solutions  are
\begin{eqnarray}
x(t)&=&x(0)+ A_{1}(t) \pi_{x}(0)+A_{2}(t)\pi_{y}(0)
+I_x(t)+I'_x(t) ,\nonumber\\
y(t)&=&y(0)+ B_{1}(t) \pi_{y}(0)-B_{2}(t) \pi_{x}(0)
- I_y(t)+I'_y(t), \nonumber\\
\pi_x(t)&=& C_{1}(t) \pi_{x}(0)+C_{2}(t)\pi_{y}(0)
+I_{\pi_x}(t)+I'_{\pi_x}(t), \nonumber\\
\pi_y(t)&=&D_{1}(t)\pi_{y}(0)-D_{2}(t)\pi_{x}(0)
-I_{\pi_y}(t)+I'_{\pi_y}(t),
\label{equ17}
\end{eqnarray}
where
$$I_x(t)=\int_0^t A_{1}(\tau)F_{x}(t-\tau)d\tau, \hspace{0.3in}
I'_x(t)=\int_0^t A_{2}(\tau)F_{y}(t-\tau)d\tau,$$
$$I_y(t)=\int_0^t
B_{2}(\tau)F_{x}(t-\tau)d\tau, \hspace{0.3in}   I'_y(t)=\int_0^t
B_{1}(\tau)F_{y}(t-\tau)d\tau,$$
$$I_{\pi_x}(t)=\int_0^t
C_{1}(\tau)F_{x}(t-\tau)d\tau, \hspace{0.3in} I'_{\pi_x}(t)=\int_0^t
C_{2}(\tau)F_{y}(t-\tau)d\tau,$$
$$I_{\pi_y}(t)=\int_0^t
D_{2}(\tau)F_{x}(t-\tau)d\tau, \hspace{0.3in} I'_{\pi_y}(t)=\int_0^t
D_{1}(\tau)F_{y}(t-\tau)d\tau,$$
and the following time-dependent
coefficients:
\begin{eqnarray}
A_{1}(t)&=&\dot{A}_{3}(t),\hspace{.3in}
A_{2}(t)=\dot{B}_{3}(t)|_{x\leftrightarrow y},\nonumber\\
A_{3}(t)&=&\frac{1}{m_x}(\frac{\lambda_y}{\lambda_x\lambda_y+\omega_{c}^2}t+\frac{\omega_{c}^2(\gamma-\lambda_y)-
\lambda_y^2(\gamma-\lambda_x)}{\gamma(\lambda_x\lambda_y+\omega_{cx}\omega_{cy})^2}\nonumber\\
&+&\sum_{i=1}^4 \frac{b_{i}e^{s_it}(\gamma+s_{i})(\gamma\lambda_y+s_{i}(\gamma+s_{i}))}{s_{i}^{2}}),\nonumber\\
B_{1}(t)&=&\dot{A}_3(t)|_{x\leftrightarrow y},\hspace{.3in}
B_2(t)=\dot{B}_{3}(t),\nonumber\\
B_{3}(t)&=&\frac{\omega_{cx}}{m_y}\left(\frac{t}{\lambda_x\lambda_y+\omega_{cx}\omega_{cy}}+\frac{2
\lambda_x\lambda_y-\gamma(\lambda_x+\lambda_y)}{\gamma(\lambda_x\lambda_y+\omega_{cx}\omega_{cy})^2}
+\sum_{i=1}^4 \frac{b_{i}e^{s_it}(\gamma+s_{i})^{2}}{s_{i}^{2}}\right),\nonumber\\
C_{1}(t)&=&m_x\ddot{A}_{3}(t),\hspace{.3in} C_{2}(t)=m_x
\ddot{B}_{3}(t),\hspace{.3in}
C_3(t)=m_x\dot{A}_3(t),\nonumber\\
D_1(t)&=&C_1(t)|_{x\leftrightarrow y},\hspace{.3in} D_2(t)=m_y
\ddot{B}_{3}(t),\hspace{.3in} D_3(t)=m_y\dot{B}_{3}(t).
\label{equ22}
\end{eqnarray}
Here,  we assume that there is no correlation  between $F_x^\nu$
and $F_y^\nu$, so that $K_{xy}=K_{yx}=0$, and
$b_i=[\prod_{j\neq i}(s_i-s_j)]^{-1}$ with $i,j=1,2,3,4$ and
$s_i$ are the roots of the  equation
\begin{eqnarray}
\gamma\lambda_x
[\gamma\lambda_y+s(\gamma+s)]+(\gamma+s)(s[s^{2}+\omega_{c}^2]
+\gamma[\omega_{c}^2+s(\lambda_y+s)])=0.
\label{equ23}
\end{eqnarray}
We introduce the spectral density $D_{\omega}$
of the  heat bath excitations  to replace the sum
over different oscillators, $\nu$, by an integral over the
frequency: $\sum_{\nu}^{}...\to \int\limits_{0}^{\infty}d\omega
D_{\omega}...$. This   is accompanied by the following
replacements: $\alpha_\nu\to\alpha_{\omega}$, $\beta_\nu\to
\beta_{\omega},$ $\omega_\nu\to\omega$, and $n_\nu\to n_{\omega}$.
Let us consider the following spectral functions \cite{Katia}
\begin{eqnarray}
D_{\omega}\frac{|\alpha_{\omega}|^2}{\omega}=
\frac{\lambda_x^2}{\pi}\frac{\gamma^2}{\gamma^2+\omega^2},\hspace{.3in}
D_{\omega}\frac{|\beta_{\omega}|^2}{\omega}=
\frac{\lambda_y^2}{\pi}\frac{\gamma^2}{\gamma^2+\omega^2},
\label{equ20}
\end{eqnarray}
where the memory time $\gamma^{-1}$ of the dissipation  is inverse
to the phonon bandwidth of the heat bath excitations which are
coupled to a quantum particle and
the
coefficients
$$\lambda_x=\hbar \alpha^2=\frac{1}{m_x}\int_0^\infty K_{xx}(t-\tau)d \tau, \hspace{.3in}
\lambda_y=\hbar \beta^2=\frac{1}{m_y}\int_0^\infty K_{yy}(t-\tau)d
\tau$$
 are the friction
coefficients in the Markovian limit.
This is  the Ohmic
dissipation with the Lorentian cutoff (Drude dissipation)
\cite{Kampen1,Kampen2,Kampen3,Kampen4,Kampen5,Kampen6,Katia,Kanokov,Kanokov2} with the dissipative
kernels
\begin{eqnarray}
K_{xx}(t)&=&m_x \lambda_x\gamma e^{-\gamma|t|}, \hspace{.3in}
K_{yy}(t)=m_y \lambda_y\gamma e^{-\gamma|t|}.
\label{equ21}
\end{eqnarray}

\subsection{Derivation of non-stationary transport coefficients}
In order to determine the transport coefficients, we use Eqs.
(\ref{equ17}). Averaging them over the whole system and by
differentiating in $t$, we obtain a  system of equations for
the first moments:
\begin{eqnarray}
<\dot{x}(t)>&=&\frac{<\pi_x(t)>}{m_x},\hspace{.3in}
<\dot{y}(t)>=\frac{<\pi_y(t)>}{m_y},\nonumber\\
<\dot{\pi}_x(t)>&=&\tilde{\omega}_{cy}(t)<\pi_y(t)>-\lambda_{\pi_x}(t)<\pi_x(t)>
,\nonumber\\
<\dot{\pi}_y(t)>&=&-\tilde{\omega}_{cx}(t)<\pi_x(t)>-\lambda_{\pi_y}(t)<\pi_y(t)>
,
\label{equ24}
\end{eqnarray}
where the friction coefficients are
\begin{eqnarray}
\lambda_{\pi_{x}}(t)=-\frac{D_{1}(t)\dot{C}_{1}(t)+D_{2}(t)\dot{C}_{2}(t)}{C_{1}(t)D_{1}(t)+C_{2}(t)D_{2}(t)},
\nonumber\\
\lambda_{\pi_{y}}(t)=-\frac{C_{1}(t)\dot{D}_{1}(t)+C_{2}(t)\dot{D}_{2}(t)}{C_{1}(t)D_{1}(t)+C_{2}(t)D_{2}(t)},
\end{eqnarray}
and the renormalized cyclotron frequencies are given by
\begin{eqnarray}
\tilde{\omega}_{cx}(t)=\frac{D_{1}(t)\dot{D}_{2}(t)-D_{2}(t)\dot{D}_{1}(t)}{C_{1}(t)D_{1}(t)+C_{2}(t)D_{2}(t)},
\nonumber\\
\tilde{\omega}_{cy}(t)=\frac{C_{1}(t)\dot{C}_{2}(t)-C_{2}(t)\dot{C}_{1}(t)}{C_{1}(t)D_{1}(t)+C_{2}(t)D_{2}(t)}.
\label{equ25}
\end{eqnarray}
As seen,
the dynamics is governed by the non-stationary coefficients.

The equations for the second moments (variances),
$$\Sigma_{q_iq_j}(t)=\frac{1}{2}<q_i(t)q_j(t)+q_j(t)q_i(t)>-<q_i(t)><q_j(t)>,$$
where $q_i=x,y,\pi_x$, or $\pi_y$ ($i$=1-4), are
\begin{eqnarray}
\dot{\Sigma}_{xx}(t)&=&\frac{2\Sigma_{x\pi_x}(t)}{m_x},\hspace{.3in}
\dot{\Sigma}_{yy}(t)=\frac{2\Sigma_{y\pi_y}(t)}{m_y},\nonumber\\
\dot{\Sigma}_{xy}(t)&=&\frac{\Sigma_{x\pi_y}(t)}{m_y}+\frac{\Sigma_{y\pi_x}(t)}{m_x},\nonumber\\
\dot{\Sigma}_{x\pi_y}(t)&=&-\lambda_{\pi_y}(t)\Sigma_{x\pi_y}(t)-\tilde{\omega}_{cx}(t)\Sigma_{x\pi_x}(t)+\frac{\Sigma_{\pi_x\pi_y}(t)}{m_x}+2D_{x\pi_y}(t),\nonumber\\
\dot{\Sigma}_{x\pi_x}(t)&=&-\lambda_{\pi_x}(t)\Sigma_{x\pi_x}(t)+\tilde{\omega}_{cy}(t)\Sigma_{x\pi_y}(t)+\frac{\Sigma_{\pi_x\pi_x}(t)}{m_x}+2D_{x\pi_x}(t),\nonumber\\
\dot{\Sigma}_{y\pi_x}(t)&=&-\lambda_{\pi_x}(t)\Sigma_{y\pi_x}(t)+\tilde{\omega}_{cy}(t)\Sigma_{y\pi_y}(t)+\frac{\Sigma_{\pi_x\pi_y}(t)}{m_y}+2D_{y\pi_x}(t),\nonumber\\
\dot{\Sigma}_{y\pi_y}(t)&=&-\lambda_{\pi_y}(t)\Sigma_{y\pi_y}(t)-\tilde{\omega}_{cx}(t)\Sigma_{y\pi_x}(t)+\frac{\Sigma_{\pi_y\pi_y}(t)}{m_y}+2D_{y\pi_y}(t),\nonumber\\
\dot{\Sigma}_{\pi_y\pi_y}(t)&=&-2\lambda_{\pi_y}(t)\Sigma_{\pi_y\pi_y}(t)-2\tilde{\omega}_{cx}(t)\Sigma_{\pi_x\pi_y}(t)+2D_{\pi_y\pi_y}(t),\nonumber\\
\dot{\Sigma}_{\pi_x\pi_x}(t)&=&-2\lambda_{\pi_x}(t)\Sigma_{\pi_x\pi_x}(t)+2\tilde{\omega}_{cy}(t)\Sigma_{\pi_x\pi_y}(t)+2D_{\pi_x\pi_x}(t),\nonumber\\
\dot{\Sigma}_{\pi_x\pi_y}(t)&=&-(\lambda_{\pi_x}(t)+\lambda_{\pi_y}(t))\Sigma_{\pi_x\pi_y}(t)+
\tilde{\omega}_{cy}(t)\Sigma_{\pi_y\pi_y}(t)-\tilde{\omega}_{cx}(t)\Sigma_{\pi_x\pi_x}(t)
+2D_{\pi_x\pi_y}(t).
\label{equ26}
\end{eqnarray}
So, we have obtained the  local in time  equations
for the first and second moments, but with the transport
coefficients depending explicitly on time. The time-dependent
diffusion coefficients $D_{q_iq_j}(t)$ are determined as
\begin{eqnarray}
D_{xx}(t)&=&D_{yy}(t)=D_{xy}(t)=0,\nonumber\\
D_{\pi_x\pi_x}(t)&=&\lambda_{\pi_x}(t)J_{\pi_x\pi_x}(t)-\tilde{\omega}_{cy}(t)J_{\pi_x\pi_y}(t)+\frac{1}{2}\dot{J}_{\pi_x\pi_x}(t),\nonumber\\
D_{\pi_y\pi_y}(t)&=&\lambda_{\pi_y}(t)J_{\pi_y\pi_y}(t)+\tilde{\omega}_{cx}(t)J_{\pi_x\pi_y}(t)+\frac{1}{2}\dot{J}_{\pi_y\pi_y}(t),\nonumber\\
D_{\pi_x\pi_y}(t)&=&-\frac{1}{2}\left[-(\lambda_{\pi_x}(t)+\lambda_{\pi_y}(t))J_{\pi_x\pi_y}(t)+
\tilde{\omega}_{cy}(t)J_{\pi_y\pi_y}(t)-\tilde{\omega}_{cx}(t)J_{\pi_x\pi_x}(t)-\dot{J}_{\pi_x\pi_y}(t)\right],\nonumber\\
D_{x\pi_y}(t)&=&-\frac{1}{2}\left[-\lambda_{\pi_y}(t)J_{x\pi_y}(t)-\tilde{\omega}_{cx}(t)J_{x\pi_x}(t)+\frac{J_{\pi_x\pi_y}(t)}{m_x}-\dot{J}_{x\pi_y}(t)\right],\nonumber\\
D_{y\pi_x}(t)&=&-\frac{1}{2}\left[-\lambda_{\pi_x}(t)J_{y\pi_x}(t)+\tilde{\omega}_{cy}(t)J_{y\pi_y}(t)+\frac{J_{\pi_x\pi_y}(t)}{m_y}-\dot{J}_{y\pi_x}(t)\right],\nonumber\\
D_{x\pi_x}(t)&=&-\frac{1}{2}\left[-\lambda_{\pi_x}(t)J_{x\pi_x}(t)+\tilde{\omega}_{cy}(t)J_{x\pi_y}(t)+\frac{J_{\pi_x\pi_x}(t)}{m_x}-\dot{J}_{x\pi_x}(t)\right],\nonumber\\
D_{y\pi_y}(t)&=&-\frac{1}{2}\left[-\lambda_{\pi_y}(t)J_{y\pi_y}(t)-\tilde{\omega}_{cx}(t)J_{y\pi_x}(t)+\frac{J_{\pi_y\pi_y}(t)}{m_y}-\dot{J}_{y\pi_y}(t)]\right].
\label{equ27}
\end{eqnarray}
Here, $\dot{J}_{q_iq_j}(t)=dJ_{q_iq_j}(t) / dt$ and the explicit expressions for $J_{q_iq_j}(t)$ are given in Appendix
A. In our treatment
$D_{xx}=D_{yy}=D_{xy}=0$ because there are no random
forces for   $x$ and $y$ coordinates in Eqs.~(\ref{equ8}). If
$\omega_{cx}=\omega_{cy}=0$, then $D_{y \pi_x}(t)=D_{x
\pi_y}(t)=D_{\pi_x \pi_y}(t)=0$.

\subsection{Asymptotic cyclotron frequency and friction  coefficients}
Using the
relationship $s_1s_2s_3s_4=\gamma^2(\lambda_{x}\lambda_{y}+\omega_{c}^2)$
between the roots of Eq.~(\ref{equ23}),  we
obtain the asymptotic ($t\rightarrow\infty$) expressions   for the friction coefficients
\begin{eqnarray}
\lambda_{\pi_{x}}(\infty)=-\frac{[\gamma+s_1+s_2]
[\gamma \lambda_y+\omega_{c}^2+(s_1+\gamma)(s_1+s_2)+s_2^2]}{(\gamma+s_1+s_2)^2+\omega_{c}^2},\nonumber\\
\lambda_{\pi_{y}}(\infty)=-\frac{[\gamma+s_1+s_2][\gamma
\lambda_x+\omega_{c}^2+(s_1+\gamma)(s_1+s_2)+s_2^2]}{(\gamma+s_1+s_2)^2+\omega_{c}^2},
\label{equ31}
\end{eqnarray}
  renormalized frequencies
\begin{eqnarray}
\tilde{\omega}_{cx}(\infty)=\frac{\omega_{cx}[(s_1+\gamma)(s_2+\gamma)-\gamma
\lambda_x]}{(\gamma+s_1+s_2)^2+\omega_{c}^2},\nonumber\\
\tilde{\omega}_{cy}(\infty)=\frac{\omega_{cy}[(s_1+\gamma)(s_2+\gamma)-\gamma
\lambda_y]}{(\gamma+s_1+s_2)^2+\omega_{c}^2},
\label{equ31}
\end{eqnarray}
and  renormalized cyclotron frequency
\begin{eqnarray}
\tilde{\omega}_{c}(\infty)=
\sqrt{\tilde{\omega}_{cx}\tilde{\omega}_{cy}},
\label{equ32}
\end{eqnarray}
where   $s_1$ and $s_2$ are the roots with the smallest absolute
values of their real parts.
As seen from Eqs. (\ref{equ31}), $\tilde{\omega}_{cx,cy}(\infty)\to\omega_{c}$ at $\gamma\to\infty$ or $\lambda_{x,y}\to 0$.



\subsection{Asymptotic variances and diffusion  coefficients}
Taking into consideration that
$\Sigma_{q_iq_j}(\infty) =J_{q_iq_j}(\infty)$, we  find
asymptotic variances
\begin{eqnarray}
\Sigma_{\pi_x \pi_y}(\infty)&=&J_{\pi_x\pi_y}(\infty)=0, \nonumber\\
\Sigma_{\pi_x\pi_x}(\infty)&=&J_{\pi_x\pi_x}(\infty)=\frac{\hbar\gamma^2m_x}{\pi}  \nonumber\\
&\times& \int_{0}^{\infty} \frac{d\omega\coth\left[\frac{\hbar
\omega}{2T}\right][\lambda_y
\gamma^2(\lambda_x\lambda_y+\omega_c^2)\omega+(\lambda_x
\gamma[\gamma-2 \lambda_y]+\lambda_y\omega_c^2)\omega^3+\lambda_x \omega^5]}
{(\omega^2+s_1^2)(\omega^2+s_2^2)(\omega^2+s_3^2)(\omega^2+s_4^2)},\nonumber\\
\Sigma_{x\pi_x}(\infty)&=&J_{x\pi_x}(\infty)\nonumber\\
&=&\frac{\hbar \gamma^2}{\pi(\lambda_x \lambda_y+\omega_c^2)}\int_{0}^{\infty}\frac{d\omega\coth[\frac{\hbar\omega}{2T}] }{(\omega^2+\gamma^2)(\omega^2+s_1^2)(\omega^2+s_2^2)(\omega^2+s_3^2)(\omega^2+s_4^2)} \nonumber\\
&\times& \{\gamma^3\lambda_y^2\omega(\lambda_x-\gamma)(\lambda_x \lambda_y+\omega_c^2)+\lambda_x\lambda_y\omega^5([\lambda_x+2(\lambda_y-\gamma)]\gamma+\omega_c^2)-\omega^7\lambda_x
\lambda_y\nonumber\\
&-&\gamma \lambda_y\omega^3(\gamma \lambda_x[\lambda_y(\lambda_y+2\lambda_x)-\gamma(2\lambda_y+\lambda_x)+\gamma^2]+\omega_c^2[\lambda_y \lambda_x+\gamma(\lambda_y-\lambda_x)])\},\nonumber\\
\Sigma_{x\pi_y}(\infty)&=&J_{x\pi_y}(\infty)\nonumber\\
&=&-\frac{\hbar \gamma^2 \omega_{cx}}{\pi(\lambda_x \lambda_y+\omega_c^2)}\int_{0}^{\infty}\frac{d\omega\coth\left[\frac{\hbar \omega}{2T}\right]}{(\omega^2+\gamma^2)(\omega^2+s_1^2)(\omega^2+s_2^2)(\omega^2+s_3^2)(\omega^2+s_4^2)}\nonumber\\
&\times& \{\gamma^3\lambda_x\omega(\gamma-\lambda_y)(\lambda_x \lambda_y+\omega_c^2)+\gamma\omega^3(\omega_c^2 \lambda_x[2 \gamma+\lambda_y]-\lambda_y[2 \lambda_x(\lambda_x \lambda_y+\omega_c^2)\nonumber\\
&-&\lambda_x \gamma(3 \lambda_y+2[\lambda_x-\gamma])-\gamma^2(\gamma-\lambda_y)])\nonumber\\
&+&\omega^5(\lambda_x
\omega_c^2+\lambda_y[\lambda_x(\lambda_x+\lambda_y)+2
\gamma(\gamma-\lambda_x)-\lambda_y \gamma]) +\omega^7\lambda_y
\},\nonumber\\
\Sigma_{\pi_y\pi_y}(\infty)&=&\Sigma_{\pi_x\pi_x}(\infty)|_{x\leftrightarrow
y},\hspace{.05 in}
\Sigma_{y\pi_y}(\infty)=\Sigma_{x\pi_x}(\infty)|_{x\leftrightarrow
y},\hspace{.05 in}
\Sigma_{y\pi_x}(\infty)=-\Sigma_{x\pi_y}(\infty)|_{x\leftrightarrow
y}.
\label{equ33}
\end{eqnarray}
The explicit expressions of asymptotic variances
at low and high temperature limits are given in Appendix B.

At $t\rightarrow\infty$, the system reaches the quasi-equilibrium state.
Taking zeros in the left parts of
Eqs.~(\ref{equ26}) for $\Sigma_{\pi_x\pi_x}$,
$\Sigma_{\pi_y\pi_y}$, $\Sigma_{\pi_x\pi_y}$, $\Sigma_{x\pi_x}$, $\Sigma_{x\pi_y}$,
$\Sigma_{y\pi_y}$, $\Sigma_{y\pi_x}$,
we obtain a linear system of
equations which establishes the one-to-one correspondence between
the asymptotic variances and asymptotic diffusion coefficients:
\begin{eqnarray}
D_{\pi_x\pi_x}(\infty)&=&\lambda_{\pi_x}(\infty)\Sigma_{\pi_x\pi_x}(\infty),\hspace{.3in}
D_{\pi_y\pi_y}(\infty)=\lambda_{\pi_y}(\infty)\Sigma_{\pi_y\pi_y}(\infty),\nonumber\\
D_{\pi_x\pi_y}(\infty)&=&\frac{1}{2}\left[\tilde{\omega}_{cx}(\infty)\Sigma_{\pi_x\pi_x}(\infty)-\tilde{\omega}_{cy}(\infty)\Sigma_{\pi_y\pi_y}(\infty)\right],\nonumber\\
D_{x\pi_y}(\infty)&=&\frac{1}{2}\lambda_{\pi_y}(\infty)\Sigma_{x\pi_y}(\infty),\hspace{.3in}
D_{y\pi_x}(\infty)=\frac{1}{2}\lambda_{\pi_x}(\infty)\Sigma_{y\pi_x}(\infty),\nonumber\\
D_{x\pi_x}(\infty)&=&-\frac{1}{2}\left[\tilde{\omega}_{cy}(\infty)\Sigma_{x\pi_y}(\infty)+\frac{1}{m_x}\Sigma_{\pi_x\pi_x}(\infty)\right],\nonumber\\
D_{y\pi_y}(\infty)&=&\frac{1}{2}\left[\tilde{\omega}_{cx}(\infty)\Sigma_{y\pi_x}(\infty)-\frac{1}{m_y}\Sigma_{\pi_y\pi_y}(\infty)\right].
\label{equ37}
\end{eqnarray}
In the axial symmetric
case ($m_x= m_y$ or $\omega_{cx}= \omega_{cy}$) with
$K_{xx}(t,\tau)=K_{yy}(t,\tau)$, we have
$\tilde{\omega}_{cx}(\infty)=\tilde{\omega}_{cy}(\infty)$,
$D_{\pi_x\pi_x}(\infty)=D_{\pi_y\pi_y}(\infty)$, $
D_{x\pi_y}(\infty)=-D_{y\pi_x}(\infty)$,
$D_{\pi_x\pi_y}(\infty)=0$,
$\Sigma_{\pi_x\pi_x}(\infty)=\Sigma_{\pi_y\pi_y}(\infty)$,
$\Sigma_{xx}(\infty)=\Sigma_{yy}(\infty)$, and
$\Sigma_{x\pi_y}(\infty)=-\Sigma_{y\pi_x}(\infty)$.

\subsection{Orbital magnetic moment}
Using Eqs.~(\ref{equ17}) and (\ref{equ22}), one can find the $z$-component of the angular
momentum in the axial symmetric case ($m_x=m_y=m$ or $\omega_{cx}=\omega_{cy}=\omega_c$)
\begin{eqnarray}
\label{390}
L_z(t)&=&<x(t)\pi_y(t) - y(t)\pi_x(t)>\nonumber\\
&=&\frac{m\hbar\gamma^2}{\pi}
\int_0^{\infty}\int_0^{t}\int_0^{t} \frac{d\omega d\tau d{\tilde \tau}\omega \coth\left[\frac{\hbar\omega}{2T}\right]}{\omega^2+\gamma^2}\cos(\omega[\tau-{\tilde \tau}]) \nonumber\\
&\times& \{\lambda_x[B_2(\tau)C_1({\tilde \tau})-A_1(\tau)D_2({\tilde \tau})]+\lambda_y[A_2(\tau)D_1({\tilde \tau})-B_1(\tau)C_2({\tilde \tau})]\}
\end{eqnarray}
and related with the magnetic moment per   volume unit
\begin{eqnarray}
\label{39}
M(t)&=&\frac{n e L_z(t)}{2m} \nonumber\\
&=&\frac{2n e\hbar \omega_c \gamma^2}{\pi m(\lambda_x
\lambda_y+\omega_c^2)}\sum_i b_i s_i
[\gamma+s_i][(\lambda_x+\lambda_y)(\gamma+s_i)-2\lambda_x\lambda_y]\nonumber\\
&\times&\int_0^{\infty} \frac{d\omega\coth\left[\frac{\hbar\omega}{2T}\right]\sin\left[\frac{\omega
t}{2}\right]\{s_i(e^{s_i t}-1) \cos\left[\frac{\omega t}{2}\right]+\omega (e^{s_i t}+1)\sin\left[\frac{\omega t}{2}\right]\}}{(\omega^2+\gamma^2)(\omega^2+s_i^2)}\nonumber\\
&+&\frac{n e \hbar \omega_c
\gamma^2}{\pi m}\sum_{i,j}\frac{b_ib_j[\gamma+s_i][\gamma+s_j]^2[s_i-s_j][s_i(\lambda_x+\lambda_y)(\gamma+s_i)+2\lambda_x\lambda_y \gamma]}{s_is_j}\nonumber\\
&\times&
\int_0^{\infty}\frac{d\omega\omega\coth\left[\frac{\hbar\omega}{2T}\right]}{(\omega^2+\gamma^2)(\omega^2+s_i^2)(\omega^2+s_j^2)}\nonumber\\
&\times&\{(\omega^2+s_is_j)(1+e^{(s_i+s_j)t}-[e^{s_it}+e^{s_jt}]\cos\left[\omega
t\right])\nonumber\\
&+&\omega(s_i-s_j)(e^{s_it}-e^{s_jt})\sin\left[\omega
t\right]\},
\end{eqnarray}
where $n$ is the concentration of  charge carriers.
In the Markovian limit (high temperatures), we obtain
\begin{eqnarray}
\label{40}
M(\infty)=-\frac{n e}{m}\frac{\omega_c T}{\lambda_x
\lambda_y+\omega_c^2}.
\end{eqnarray}
In the case $\lambda_x=\lambda_y$, the similar expression is derived in Ref. \cite{Ma}.
As seen, $M(\infty) $ approaches zero  with increasing friction coefficient.
This approach is slower the larger  the cyclotron frequency is.
Note that the Bohr-Van Leeuwen theorem (there is no diamagnetism  in the classical system)
is restored in the limit of infinite damping or cyclotron frequency.

At   low temperature ($T\to 0$), the magnetic moment
\begin{eqnarray}
\label{41}
\lefteqn{M(\infty)=\frac{n e \hbar \omega_c \gamma^2}{\pi m(\lambda_x
\lambda_y+\omega_c^2)}\sum_i\frac{b_i s_i
[\gamma+s_i][(\lambda_x+\lambda_y)(\gamma+s_i)-2\lambda_x\lambda_y]\ln\left(\frac{\gamma^2}{s_i^2}\right)}{\gamma^2-s_i^2}}\nonumber\\
&+&\frac{n e \hbar \omega_c
\gamma^2}{\pi m}\sum_{i,j}\frac{b_ib_j[\gamma+s_i][\gamma+s_j]^2[s_i-s_j][s_i(\lambda_x+\lambda_y)(\gamma+s_i)+2\lambda_x\lambda_y]}{s_is_j} \nonumber\\
&\times&
 \frac{[s_i+s_j][\gamma^2-s_is_j]\ln(\gamma^2)-s_i[\gamma^2-s_j^2]\ln(s_i^2)-s_j[\gamma^2-s_i^2]\ln(s_j^2)}{[s_i+s_j][\gamma^2-s_i^2][\gamma^2-s_j^2]}
\end{eqnarray}
is also nonzero in the presence of dissipation. The orbital diamagnetism survives in the
dissipative environment.
At $\omega_c\gg \sqrt{\lambda_x \lambda_y}$, $\gamma\to\infty$, and $T\to 0$, we obtain
\begin{eqnarray}
\label{42}
L_z(\infty)=-\hbar,  \hspace{0.3 in}    M(\infty)=-\frac{n e \hbar}{2m}.
\end{eqnarray}
As seen, for large values of the cyclotron frequency,
a saturation value of the magnetization   equals one (negative) Bohr magneton.
So, in the dissipative system, we find the quantization conditions
for the orbital  angular momentum and magnetic moment.

\section{Results of calculations}
In the model  considered, one can
investigate the properties of
friction and diffusion coefficients, cyclotron frequencies,
variances, and angular momentum or magnetic moments.
In addition, one can  also
study the magnetic moment of the system.
It should be noted  that in
our model  the influence of magnetic field on the
coupling between quantum particle and heat-bath is neglected.
The impact of the
magnetic field is entered into the dissipative kernels.
%
However, there are  solids whose
resistance remains constant in the wide spectrum of magnetic
field. Their properties   can be described by
neglecting the effect of magnetic field on the coupling term.


\subsection{Transport coefficients and variances}

 The  dependencies of $\lambda_\pi$ and $\tilde{\omega}_c$
on   time are given in  Fig. 1.
The non-Markovian correction to the friction coefficient
increases with asymptotic friction coefficient (right side) and
decreases with the magnetic field (left side). The
increase of the friction and magnetic field contributes to the
rise of asymptotic  magnetic field (bottom parts of
Fig.~1).  In general, the rise of the
asymptotic friction coefficient increases the transient time of
$\lambda_{\pi_x}$ and $\tilde{\omega}_c$.
\begin{figure}[h]
    \includegraphics[width=15cm]{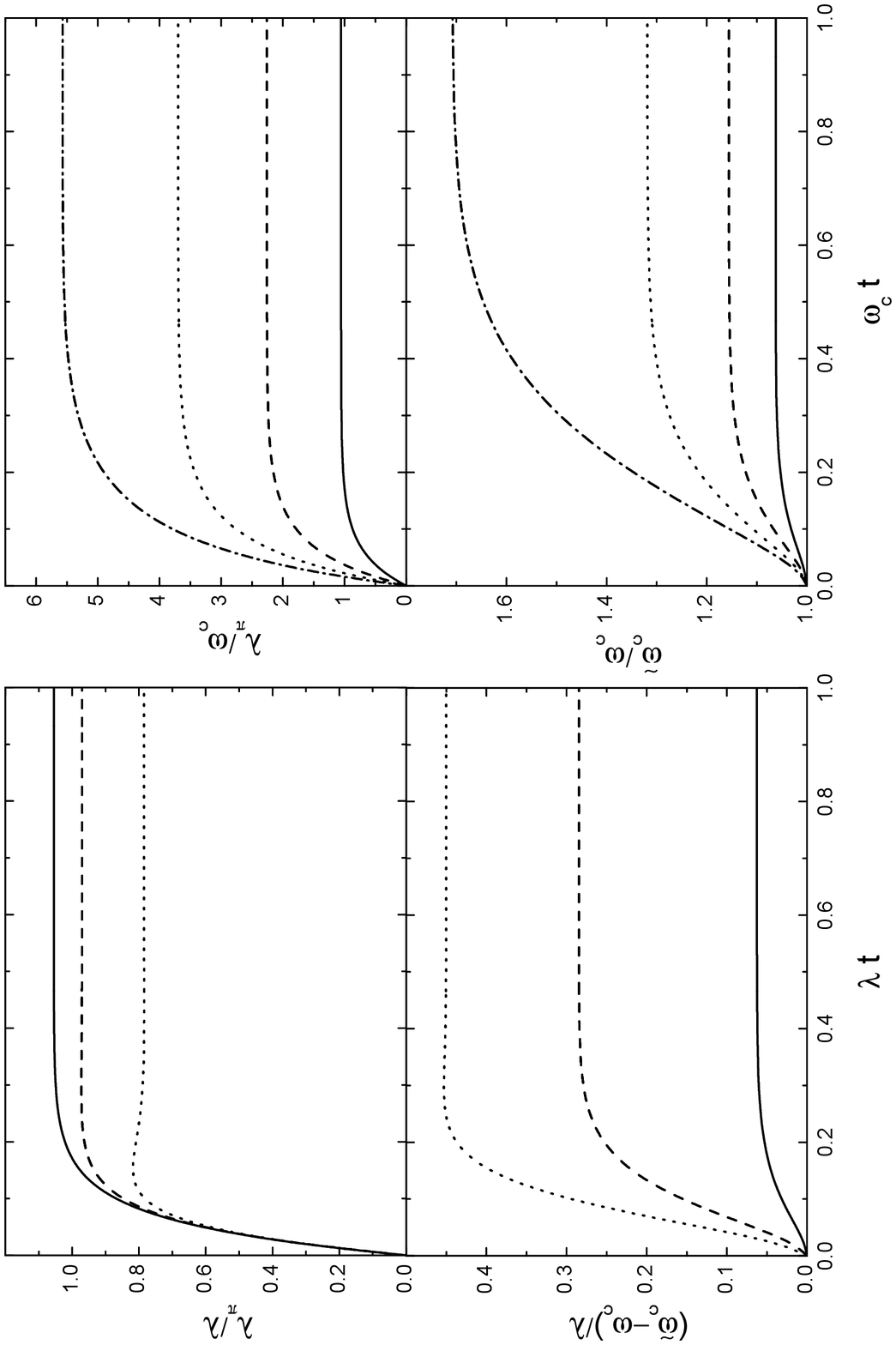}
\caption{The calculated  friction coefficient $\lambda_\pi$ and cyclotron frequency $\tilde\omega_c$  as  functions of time.
  The results for the  frequencies $\frac{\omega_c}{\lambda}$=1, 5, and 10  of the external magnetic
 field   at given Markovian friction coefficient $\lambda$
 are presented by solid, dashed, and dotted lines, respectively (left side).
 The results for the  Markovian friction coefficient  ($\lambda=\lambda_x=\lambda_y$)
 $\frac{\lambda}{\omega_c}$=1, 2,  3, and
4  at given external magnetic
 field $\omega_c$ are presented by solid, dashed, dotted, and dash-dotted lines,
respectively (right side).}
\label{1_fig}
\end{figure}
%
%
\begin{figure}[h]
  \includegraphics[width=15cm]{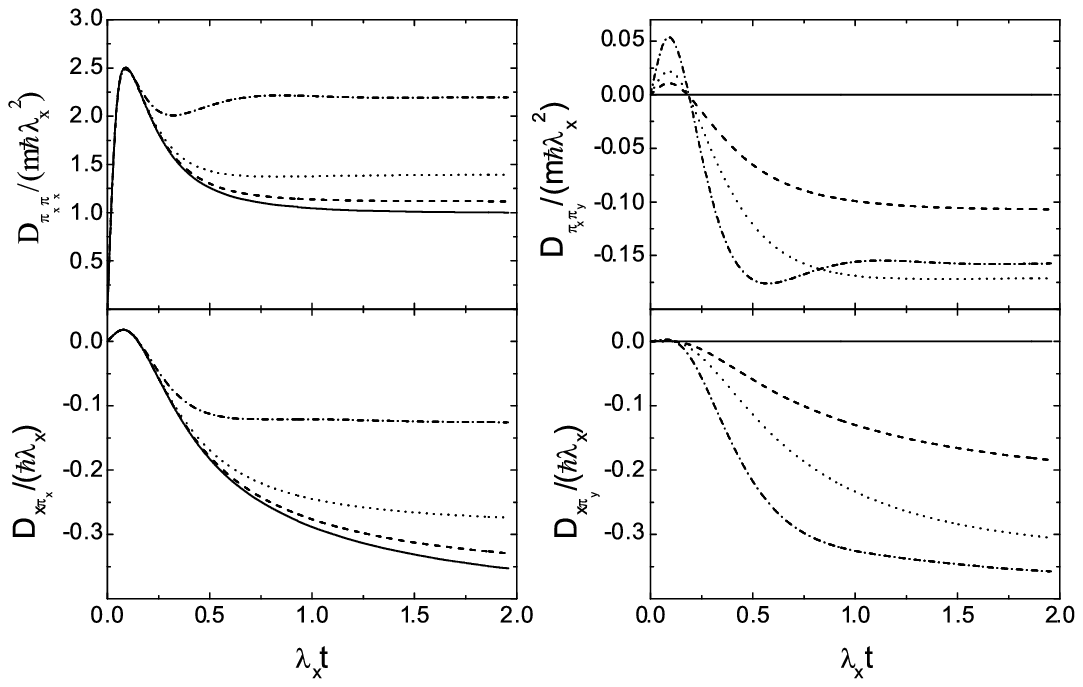}\\
  \caption{The calculated time dependence
of the diffusion coefficients $D_{\pi_x\pi_x}$, $D_{\pi_x\pi_y}$,
$D_{x\pi_x}$, and $D_{x\pi_y}$ at  low temperature $T/(\hbar
\lambda_x)=0.1$ and $\lambda_y/\lambda_x=2$.
The results for $\omega_c/\lambda_x=0$, 1, 2, and 5
are presented by solid, dashed, dotted, and dash-dotted lines,
respectively.}
\label{2_fig}
\end{figure}
\begin{figure}[p]
  \includegraphics[width=15cm]{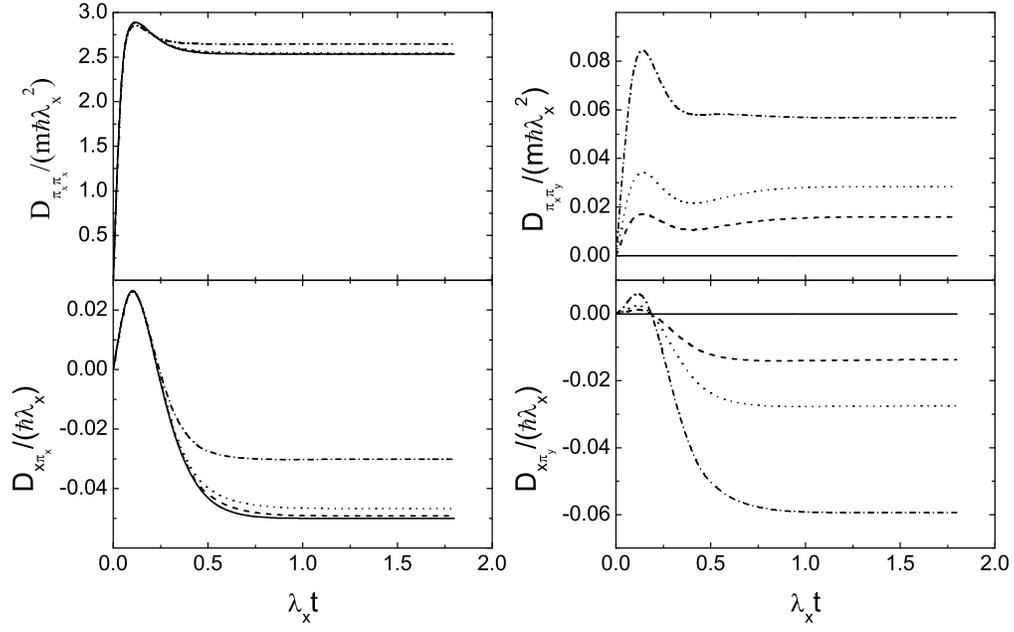}\\
  \caption{The same as in Fig. 2, but at temperature $T/(\hbar
\lambda_x)=2$.}
\label{3_fig}
\end{figure}
\begin{figure}[h!]
  \includegraphics[width=15cm]{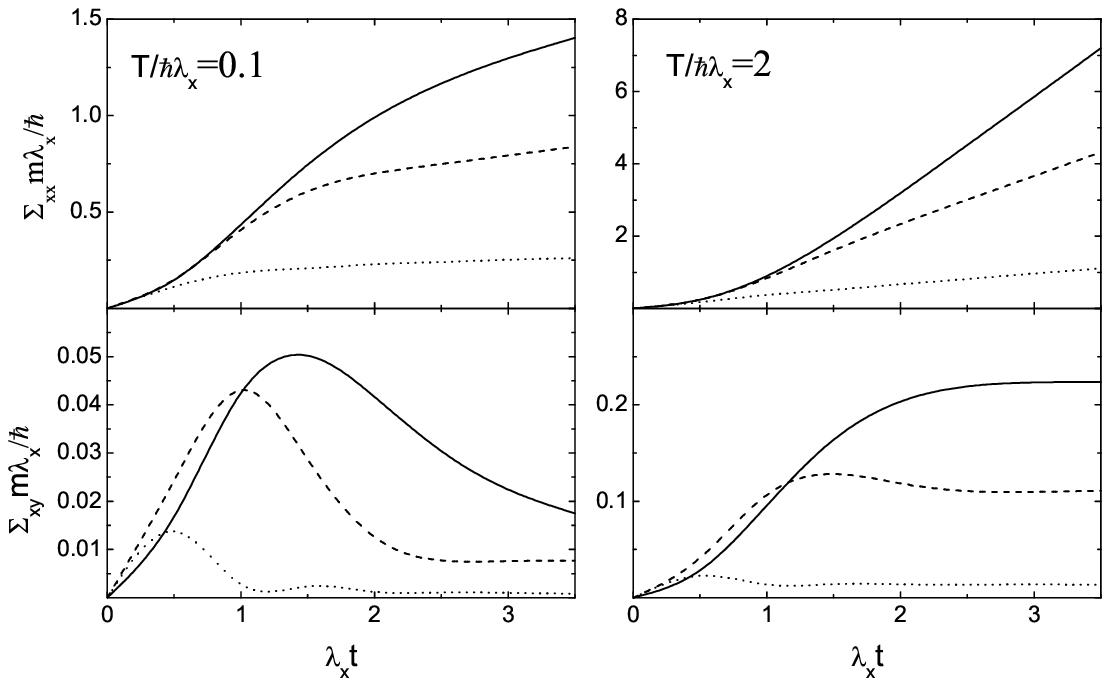}\\
  \caption{The calculated time-dependent   variances
  $\Sigma_{xx}$ and $\Sigma_{xy}$ at   indicated temperatures
  and $\lambda_y/\lambda_x=2$.
  The solid, dashed, and dotted lines
  correspond to $\omega_c/\lambda_x=1$, 2, and 3, respectively.}
  \label{4_fig}
\end{figure}

The time evolutions of the diffusion coefficients $D_{\pi_x\pi_x}$,
$D_{x\pi_x}$ , $D_{x\pi_y}$ , and $D_{\pi_x\pi_y}$ at different
temperatures are shown in  Figs.~2 and 3.
These coefficients are
initially equal to zero, and in some transient time, reach
their asymptotic values.
At   low temperature, the asymptotic
value of $D_{\pi_x\pi_x}$ changes    stronger  with the
field in comparison to the case of high temperature (compare Figs.~2 and 3).
The value of $|D_{x\pi_x}(\infty)|$ decreases
with increasing $\omega_c$ and approaches nearly zero in Fig.~2.
In the absence of  magnetic field, $D_{x\pi_y} =0$. However, upon switching   the
magnetic field $D_{x\pi_y}$ becomes non-zero with a negative asymptotic value
(Fig.~3). The asymptotic  value of $|D_{x\pi_y}|$ increases with
$\omega_c$ and decreases with increasing temperature. The value of
$D_{\pi_x\pi_y}$ is equal to zero at $\lambda_x=\lambda_y$ and
becomes negative (positive) at $\lambda_x>\lambda_y$ ($\lambda_x<\lambda_y$) because
$$D_{\pi_x\pi_y}(\infty)=
\frac{1}{2}\left[\tilde{\omega}_{cx}(\infty)\Sigma_{\pi_x\pi_x}(\infty)-
\tilde{\omega}_{cy}(\infty)\Sigma_{\pi_y\pi_y}(\infty)\right]$$
and
$\Sigma_{\pi_x\pi_x}(\infty)<\Sigma_{\pi_y\pi_y}(\infty)$ at
$\lambda_x>\lambda_y$
($\Sigma_{\pi_x\pi_x}(\infty)>\Sigma_{\pi_y\pi_y}(\infty)$ at
$\lambda_x<\lambda_y$).

 The time-dependent  variances $\Sigma_{xx}$, $\Sigma_{yy}$, and
$\Sigma_{xy}$ are presented in  Fig.~4. One can see the steadily increase of
$\Sigma_{xx}$ with   time that is quite expected for the
systems without   potential confinement of particle motion.
 The time behavior of $\Sigma_{xy}$ is more
complicated and depends on the interplay between $\lambda_x$
and $\lambda_y$. The absolute values of $\Sigma_{xx}$, $\Sigma_{yy}$, and
$\Sigma_{xy}$ decreases with increasing $\omega_c$. So, the localization
of the charged particle is enhanced by an increasing   magnetic field
and  decreasing   temperature. As seen, at $T/\lambda_x=0.1$
and $\omega_c/\lambda_x=3$ ($\lambda_y/\lambda_x=2$)
the system almost reaches the quasi-equilibrium state.
The same localization phenomenon was observed for the charged particle
in the harmonic oscillator potential in a dissipative environment \cite{Ford,Kanokov2}.



The asymptotic friction
coefficients $\lambda_{x,y}$  unexpectedly decrease  with increasing value of
$\omega_c$
in the bosonic system considered.
%
%
\begin{figure}[h!]
  \includegraphics[height=15cm]{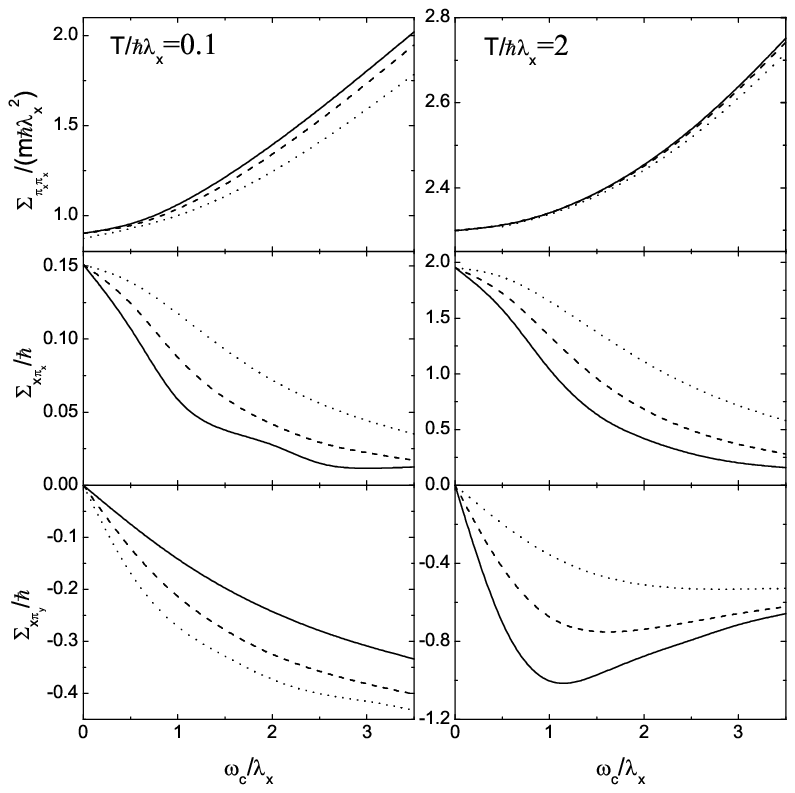}\\
  \caption{The  calculated  dependencies of the  asymptotic variances on $\omega_c$. The solid, dashed, and dotted lines
  correspond to $\lambda_y/\lambda_x=1$, $2$, and $5$, respectively. }
\label{5_fig}
\end{figure}
\begin{figure}[h!]
  \includegraphics[height=15cm]{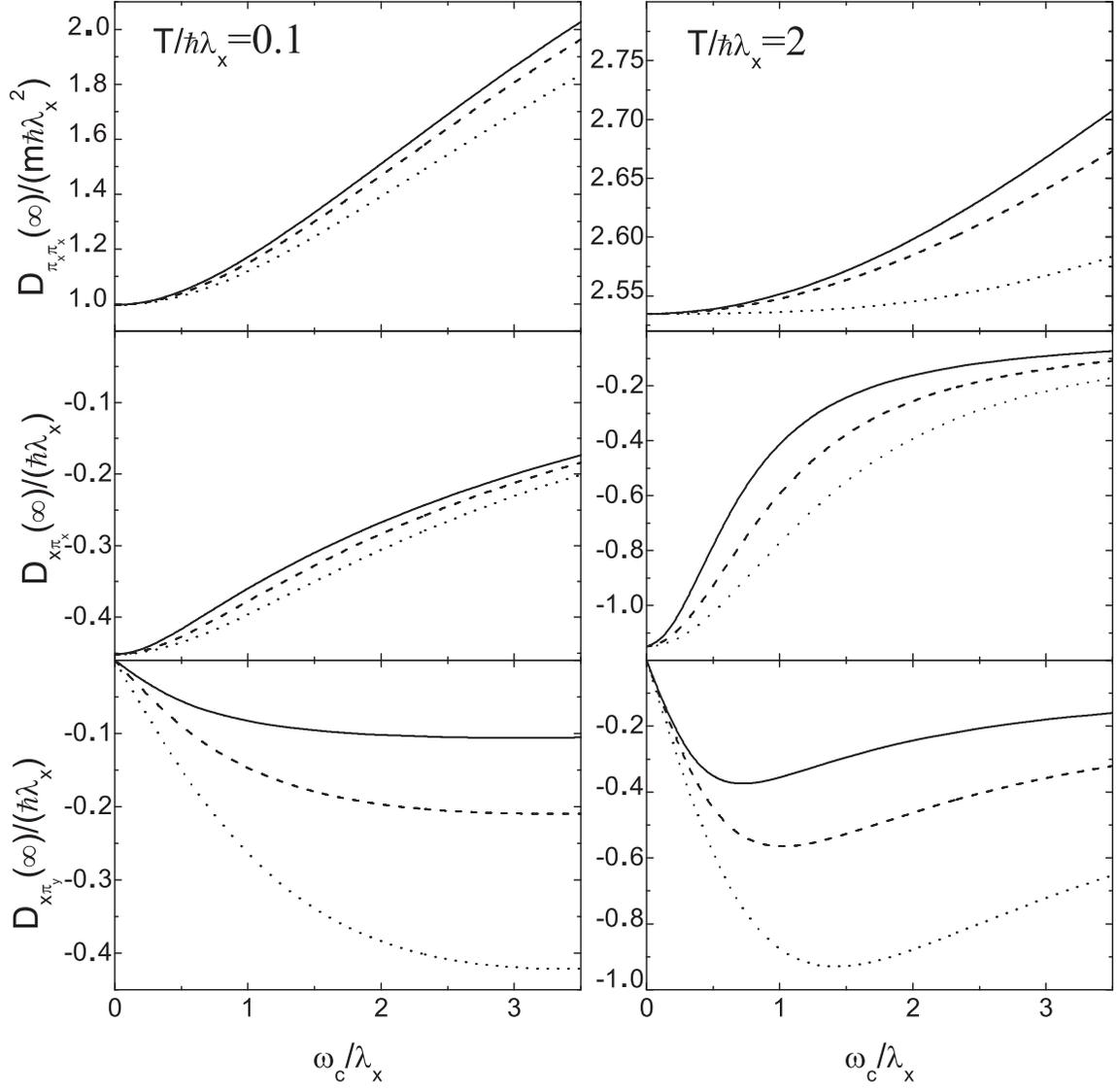}\\
  \caption{The  calculated  dependencies  of the   asymptotic diffusion coefficients on $\omega_c$. The solid, dashed, and dotted lines
  correspond to $\lambda_y/\lambda_x=0.5$, $1$, and $2$, respectively. }
\label{6_fig}
\end{figure}
\begin{figure}[h!]
  \includegraphics[height=10cm]{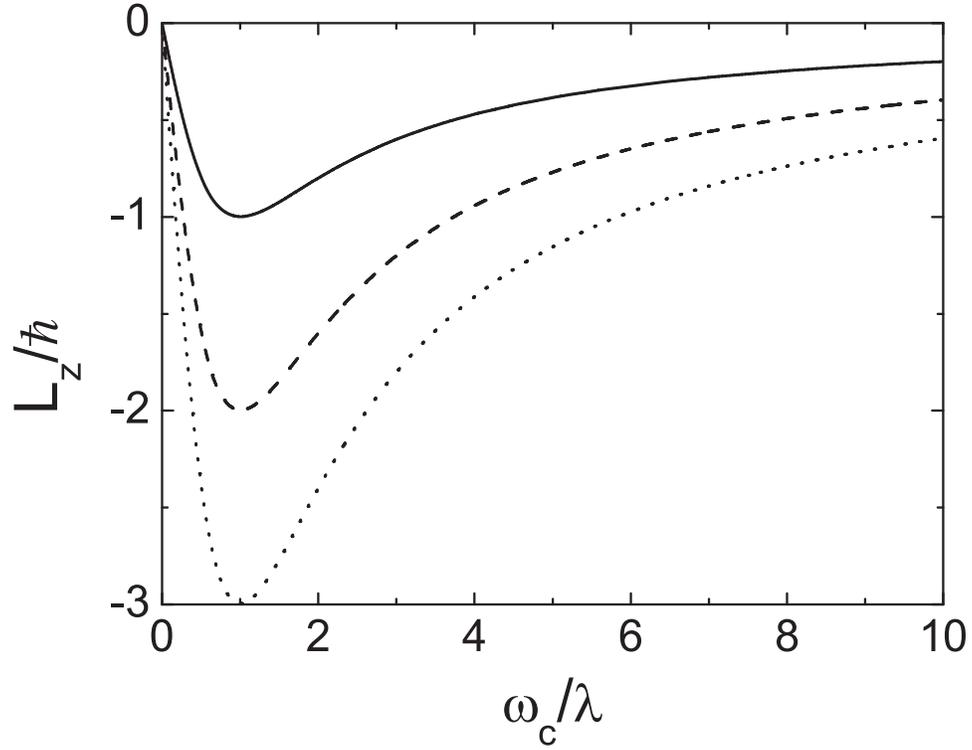}\\
  \caption{The  calculated asymptotic $z$-component of  angular momentum $L$ as
  a function of $\omega_c/\lambda$ at $\lambda_x=\lambda_y=\lambda$ and $\gamma/\lambda=12$.
  The solid, dashed, and dotted lines
  correspond to the cases with $T/(\hbar\lambda)=1$, $2$, and $3$, respectively. }
\label{7_fig}
\end{figure}
\begin{figure}[h!]
  \includegraphics[height=10cm]{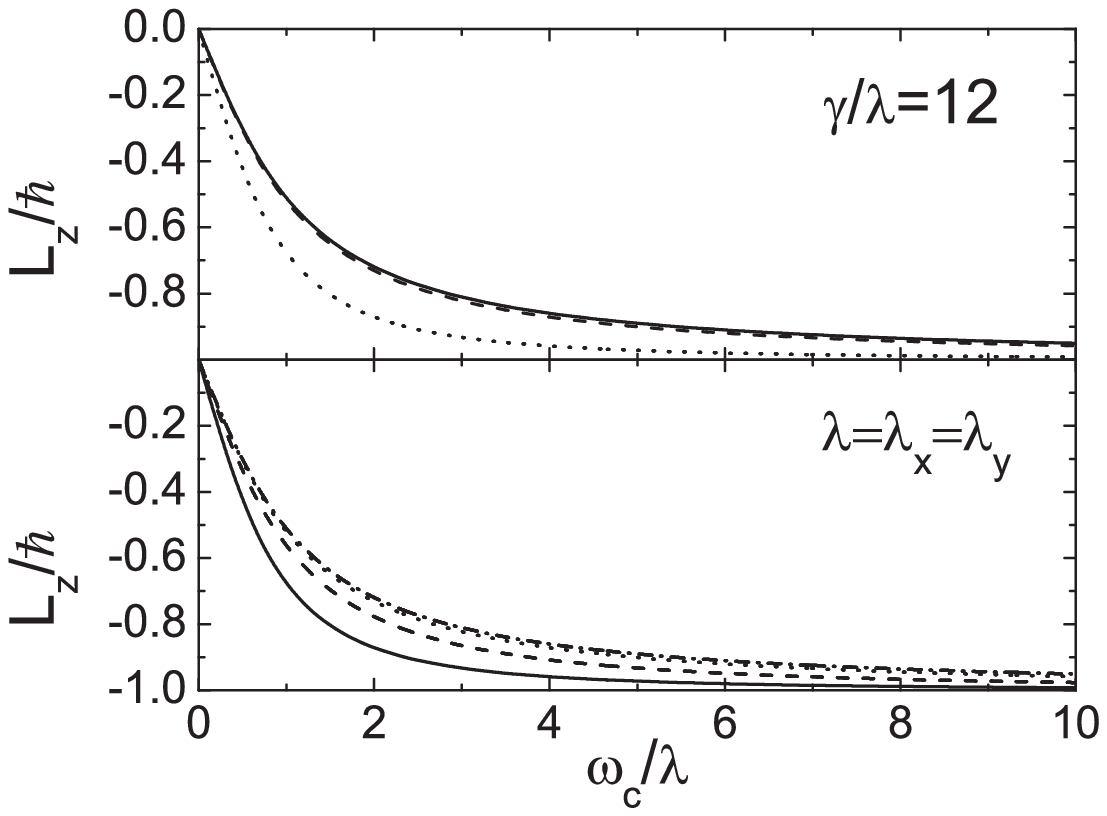}\\
  \caption{The  calculated asymptotic  $z$-component of   angular momentum $L$  as a function of $\omega_c/\lambda$ at   $T=0$.
  In upper part, $\gamma/\lambda=12$, $\lambda=\lambda_x$,    $\lambda_y=\lambda_x=$1 (solid line), 2 (dashed line), 3 (dotted line).
  In lower part,   $\lambda_y=\lambda_x=\lambda$, $\gamma/\lambda=1$ (solid line), 5 (dashed line), 20 (dotted line), 40 (dash-dotted line).
}
\label{8_fig}
\end{figure}
%
Note  that the
friction and resistance are quite different values and
the diagonal components of resistance tensor
\begin{eqnarray}
\rho(\infty)\sim
  \begin{pmatrix}
    m_x \lambda_x &&& m_x \omega_{cx} \\
    -m_y \omega_{cy} &&& m_y \lambda_y
  \end{pmatrix}\nonumber
\end{eqnarray}
 obtained in our model does not depend on  magnetic
field.
%
%
 Moreover,
the friction does not depend on   magnetic
field in the Markovian limit, $\gamma\rightarrow\infty$.
The friction
coefficients have relatively small influence on the process
  in the system   at almost all spectrum of the
magnetic field except for very weak fields. Analyzing the dependence
of the frequency of microscopic magnetic field $\tilde{\omega}_c$
on $\lambda$,
one can conclude  that
the specimen with nonzero friction perceives the external magnetic
  field  with higher intensity.
The non-Markovian corrections to  the external
magnetic   field  are larger
for the system with longer time $\gamma^{-1}$  of response.

The dependencies of asymptotic variances
and diffusion coefficients on magnetic field
are shown
in  Figs.~5 and 6.  At low temperature, the absolute values of $\Sigma_{\pi_x\pi_x}$ and
$\Sigma_{x\pi_y}$ increase with the field while
$\Sigma_{x\pi_x}$ shows the opposite trend. At high temperature, we have the same behavior for  $\Sigma_{\pi_x\pi_x}$ and $\Sigma_{x\pi_x}$,
but different dependence for $\Sigma_{x\pi_y}$. Its absolute value firstly increases with $\omega_c$, reaches the minimum, and then decreases.
As seen, at low temperature, the increase of $\lambda_y$ with respect to $\lambda_x$ leads to larger absolute values of $\Sigma_{x\pi_x}$ ($D_{x\pi_x}$) and
$\Sigma_{x\pi_y}$ ($D_{x\pi_y}$), and to smaller values of $\Sigma_{\pi_x\pi_x}$ ($D_{\pi_x\pi_x}$).
With increasing temperature,  $\Sigma_{x\pi_x}$ ($D_{x\pi_x}$) and $\Sigma_{\pi_x\pi_x}$ ($D_{\pi_x\pi_x}$) keep their
behavior unchanged.

\subsection{Orbital angular momentum component}
We calculate  the $z$-component of the angular momentum $L_z$  for the system
settled in the increasing external magnetic field at different temperatures (Figs. 7 and 8).
The results indicate the diamagnetism of the system even in the presence of a
physical heat bath.
As seen in Fig. 7, the absolute value of the magnetization of electric charges increases with temperature.
At high  $B$, the value of $L_z$ or $M$ tends to 0 as $\omega_c^{-1}$.
At low temperatures and large $\gamma$, the value of $L_z$ approaches   $\hbar$   with increasing  $B$ ( Fig. 8).
At $T\to 0$, $B\to \infty$, and $\gamma\to\infty$, we obtain the usual quantization of $L_z$
in the dissipative system.

\section{Summary}
The behavior of the  generated flow of free charge carriers under the influence of
external magnetic field was studied within   the non-Markovian two-dimensional Langevin approach
and the linear coupling in coordinate between the charge carriers and environment.
In order to average the influence of bosonic heat-bath on the
charged particle, we applied the spectral function
of   heat-bath excitations which describes the Drude dissipation with Lorenzian cutoffs.
The analytical expressions for the
time-dependent and asymptotic friction and diffusion coefficients,
variances of the  coordinates,
cyclotron frequencies,  orbital magnetic moment were obtained.
The influence of an external magnetic field on the
transport properties of an open quantum system was studied
at the limits of low and high temperatures.
Based on the calculations,
one can conclude that
the system in  dissipative environment perceives the external magnetic
  field  with higher intensity. The non-Markovian corrections to  the external
magnetic   field  are larger for the system with longer memory time.
The decrease of asymptotic friction
coefficients   and the   localization of the charged particle with
increasing magnetic field were observed for the bosonic system.
We demonstrated   the survival of diamagnetism of the system  in the presence of the
realistic heat bath at low and high temperature regimes.
%
%
For the orbital  magnetic moment or angular momentum in the dissipative system,
we  obtained the quantization condition at $T\to 0$, $B\to \infty$, and $\gamma\to\infty$.

\acknowledgments
This work was partially supported by  the Russian Foundation for Basic Research (Moscow)   and DFG (Bonn).
The IN2P3(France)-JINR(Dubna) Cooperation
Programme is gratefully acknowledged.

\appendix

\section{Coefficients $J_{q_iq_j}(t)$}
\label{sec.ap}
 The explicit expressions for the coefficients
\begin{eqnarray}
J_{xx}(t)&=&\ll I_x(t)I_x(t)+I'_x(t)I'_x(t)\gg,  \hspace{.1in}
J_{yy}(t)=J_{xx}(t)|_{x\to y},\nonumber\\
J_{xy}(t)&=&\ll I_x(t)I_y(t)+I'_x(t)I'_y(t)\gg,  \hspace{.1in} J_{\pi_x\pi_y}(t)=J_{xy}(t)|_{x\to \pi_x,y\to \pi_y}, \nonumber\\
J_{x\pi_x}(t)&=&\ll I_{x}(t)I_{\pi_x}(t)+I'_{x}(t)I'_{\pi_x}(t)\gg,  \hspace{.1in} J_{y\pi_x}(t)|=J_{x\pi_x}(t)|_{x\to y}, \nonumber\\
J_{x\pi_y}(t)&=&\ll I_{x}(t)I_{\pi_y}(t)+I'_{x}(t)I'_{\pi_y}(t)\gg, \hspace{.1in} J_{y\pi_y}(t)=J_{x\pi_y}(t)|_{x\to y},\nonumber\\
J_{\pi_x\pi_x}(t)&=&\ll I_{\pi_x}(t)I_{\pi_x}(t)+I'_{\pi_x}(t)I'_{\pi_x}(t)\gg,  \hspace{.1in}
J_{\pi_y\pi_y}(t)=J_{\pi_x\pi_x}(t)|_{pi_x\to \pi_y}.
\label{equ28}
\end{eqnarray}
are:
\begin{eqnarray}
  J_{xx}(t)&=&\frac{m \hbar \gamma^2}{\pi}\int_{0}^{\infty}d\omega\int_0^tdt'\int_0^tdt''\frac{\omega
  \coth\left[\frac{\hbar\omega}{2T}\right]}{\omega^2+\gamma^2} \nonumber\\
  &\times&\left[\lambda_xA_1(t')A_1(t'')+\lambda_yA_2(t')A_2(t'')\right]\cos(\omega[t''-t']),\nonumber
\end{eqnarray}
\begin{eqnarray}
  J_{yy}(t)&=&\frac{m \hbar \gamma^2}{\pi}\int_{0}^{\infty}d\omega\int_0^tdt'\int_0^tdt''\frac{\omega
  \coth\left[\frac{\hbar\omega}{2T}\right]}{\omega^2+\gamma^2}\nonumber\\
  &\times&\left[\lambda_xB_2(t')B_2(t'')+\lambda_yB_1(t')B_1(t'')\right]\cos(\omega[t''-t']),\nonumber
\end{eqnarray}
\begin{eqnarray}
  J_{xy}(t)&=&\frac{m \hbar \gamma^2}{\pi}\int_{0}^{\infty}d\omega\int_0^tdt'\int_0^tdt''\frac{\omega
  \coth\left[\frac{\hbar\omega}{2T}\right]}{\omega^2+\gamma^2}\nonumber\\
  &\times&\left[\lambda_xA_1(t')B_2(t'')+\lambda_yA_2(t')B_1(t'')\right]\cos(\omega[t''-t']),\nonumber
\end{eqnarray}
\begin{eqnarray}
  J_{\pi_x\pi_x}(t)&=&\frac{m \hbar \gamma^2}{\pi}\int_{0}^{\infty}d\omega\int_0^tdt'\int_0^tdt''\frac{\omega
  \coth\left[\frac{\hbar\omega}{2T}\right]}{\omega^2+\gamma^2}\nonumber\\
    &\times&\left[\lambda_xC_1(t')C_1(t'')+\lambda_yC_2(t')C_2(t'')\right]\cos(\omega[t''-t']),\nonumber
\end{eqnarray}
\begin{eqnarray}
  J_{\pi_y\pi_y}(t)&=&\frac{m \hbar \gamma^2}{\pi}\int_{0}^{\infty}d\omega\int_0^tdt'\int_0^tdt''\frac{\omega
  \coth\left[\frac{\hbar\omega}{2T}\right]}{\omega^2+\gamma^2}\nonumber\\
    &\times&\left[\lambda_xD_2(t')D_2(t'')+\lambda_yD_1(t')D_1(t'')\right]\cos(\omega[t''-t']),\nonumber
\end{eqnarray}
\begin{eqnarray}
  J_{\pi_x\pi_y}(t)&=&\frac{m \hbar \gamma^2}{\pi}\int_{0}^{\infty}d\omega\int_0^tdt'\int_0^tdt''\frac{\omega
  \coth\left[\frac{\hbar\omega}{2T}\right]}{\omega^2+\gamma^2}\nonumber\\
    &\times&\left[\lambda_xC_1(t')D_2(t'')+\lambda_yC_2(t')D_1(t'')\right]\cos(\omega[t''-t']),\nonumber
\end{eqnarray}
\begin{eqnarray}
  J_{x\pi_x}(t)&=&\frac{m \hbar \gamma^2}{\pi}\int_{0}^{\infty}d\omega\int_0^tdt'\int_0^tdt''\frac{\omega
  \coth\left[\frac{\hbar\omega}{2T}\right]}{\omega^2+\gamma^2}\nonumber\\
    &\times&\left[\lambda_xA_1(t')C_1(t'')+\lambda_yA_2(t')C_2(t'')\right]\cos(\omega[t''-t']),\nonumber
\end{eqnarray}
\begin{eqnarray}
  J_{y\pi_y}(t)&=&\frac{m \hbar \gamma^2}{\pi}\int_{0}^{\infty}d\omega\int_0^tdt'\int_0^tdt''\frac{\omega
  \coth\left[\frac{\hbar\omega}{2T}\right]}{\omega^2+\gamma^2}\nonumber\\
    &\times&\left[\lambda_xB_2(t')D_2(t'')+\lambda_yB_1(t')D_1(t'')\right]\cos(\omega[t''-t']),\nonumber
\end{eqnarray}
\begin{eqnarray}
  J_{x\pi_y}(t)&=&\frac{m \hbar \gamma^2}{\pi}\int_{0}^{\infty}d\omega\int_0^tdt'\int_0^tdt''\frac{\omega
  \coth\left[\frac{\hbar\omega}{2T}\right]}{\omega^2+\gamma^2}\nonumber\\
    &\times&\left[\lambda_xA_1(t')D_2(t'')+\lambda_yA_2(t')D_1(t'')\right]\cos(\omega[t''-t']),\nonumber
\end{eqnarray}
\begin{eqnarray}
  J_{y\pi_x}(t)&=&\frac{m \hbar \gamma^2}{\pi}\int_{0}^{\infty}d\omega\int_0^tdt'\int_0^tdt''\frac{\omega
  \coth\left[\frac{\hbar\omega}{2T}\right]}{\omega^2+\gamma^2}\nonumber\\
    &\times&\left[\lambda_xB_2(t')C_1(t'')+\lambda_yB_1(t')C_2(t'')\right]\cos(\omega[t''-t']).\nonumber
\end{eqnarray}

\section{Asymptotic variances at  high and low temperature limits}
\label{sec.ap2}
At  high temperature limit,
the asymptotic variances are
\begin{eqnarray}
&&\Sigma_{\pi_x \pi_y}(\infty)=0,\nonumber\\
&& {}\Sigma_{\pi_x\pi_x}(\infty)=m_x T,\hspace{.2 in}
\Sigma_{x\pi_x}(\infty)=\frac{\lambda_y T}{\lambda_x
\lambda_y+\omega_c^2},\hspace{.2 in}
\Sigma_{x\pi_y}(\infty)=-\frac{\omega_{cx}
T}{\lambda_x \lambda_y+\omega_c^2},\nonumber\\
&&
{}\Sigma_{\pi_y\pi_y}(\infty)=\Sigma_{\pi_x\pi_x}(\infty)|_{x\leftrightarrow
y},\hspace{.05 in}
\Sigma_{y\pi_y}(\infty)=\Sigma_{x\pi_x}(\infty)|_{x\leftrightarrow
y},\hspace{.05 in}
\Sigma_{y\pi_x}(\infty)=-\Sigma_{x\pi_y}(\infty)|_{x\leftrightarrow
y}.
\label{equ34}
\end{eqnarray}
At   low temperature limit,
we have
\begin{eqnarray}
\Sigma_{\pi_x \pi_y}(\infty)&=&0,\nonumber\\
\Sigma_{\pi_x\pi_x}(\infty)&=&\frac{\hbar \gamma^2 m }{\pi
\delta}\{\lambda_y
\gamma^2\varpi_1(\lambda_x\lambda_y+\omega_c^2)+\varpi_2(\lambda_x
\gamma[\gamma-2 \lambda_y]+\lambda_y \omega_c^2)+\varpi_3\lambda_x
\},\nonumber\\
\Sigma_{x\pi_x}(\infty)&=&\frac{\hbar \gamma^2}{\pi
(\lambda_x \lambda_y+\omega_c^2)\Delta}
\{\gamma^3\lambda_y^2\zeta_1(\lambda_x-\gamma)(\lambda_x
\lambda_y+\omega_c^2)\nonumber\\
&-&\gamma \lambda_y\zeta_2(\gamma \lambda_x[\lambda_y(\lambda_y+2\lambda_x)-
\gamma(2\lambda_y+\lambda_x)+\gamma^2]+\omega_c^2[\lambda_y \lambda_x+\gamma(\lambda_y-\lambda_x)])\nonumber\\
&+&\lambda_x
\lambda_y\zeta_3([\lambda_x+2(\lambda_y-\gamma)]\gamma+\omega_c^2)-\lambda_x
\zeta_4\lambda_y \}, \nonumber\\
\Sigma_{x\pi_y}(\infty)&=&-\frac{\hbar \gamma^2 \omega_c}{\pi(\lambda_x \lambda_y+\omega_c^2)\Delta}\{\gamma^3\lambda_x\zeta_1(\gamma-\lambda_y)(\lambda_x \lambda_y+\omega_c^2)\nonumber\\
&+&\gamma\zeta_2(\omega_c^2 \lambda_x[2 \gamma+\lambda_y]-\lambda_y[2
\lambda_x(\lambda_x \lambda_y+\omega_c^2)-
\lambda_x \gamma(3 \lambda_y+2[\lambda_x-\gamma])-\gamma^2(\gamma-\lambda_y)])\nonumber\\
&+&\zeta_3(\lambda_x
\omega_c^2+\lambda_y[\lambda_x(\lambda_x+\lambda_y)+2
\gamma(\gamma-\lambda_x)-\lambda_y \gamma]) + \zeta_4\lambda_y
\},\nonumber\\
\Sigma_{\pi_y\pi_y}(\infty)&=&\Sigma_{\pi_x\pi_x}(\infty)|_{x\leftrightarrow
y},\hspace{.05 in}
\Sigma_{y\pi_y}(\infty)=\Sigma_{x\pi_x}(\infty)|_{x\leftrightarrow
y},\hspace{.05 in}
\Sigma_{y\pi_x}(\infty)=-\Sigma_{x\pi_y}(\infty)|_{x\leftrightarrow
y},
\label{equ35}
\end{eqnarray}
where
\begin{eqnarray}
  \delta&=&(s_1^2-s_2^2)(s_1^2-s_3^2)(s_1^2-s_4^2)(s_2^2-s_3^2)(s_2^2-s_4^2)(s_3^2-s_4^2),\nonumber\\
  \Delta&=&(\gamma^2-s_1^2)(\gamma^2-s_2^2)(\gamma^2-s_3^2)(\gamma^2-s_4^2)\delta,
\label{equ36}
\end{eqnarray}
 and
\begin{eqnarray}
  \varpi_1&=&(s_1 s_2)^4(s_1^2-s_2^2)\ln\left[\frac{s_1}{s_2}\right]+(s_1 s_3)^4(s_3^2-s_1^2)\ln\left[\frac{s_3}{s_1}\right]
  +(s_1 s_4)^4(s_1^2-s_4^2)\ln\left[\frac{s_1}{s_4}\right]\nonumber\\
  &+&(s_2 s_3)^4(s_2^2-s_3^2)\ln\left[\frac{s_2}{s_3}\right]+
  (s_2 s_4)^4(s_4^2-s_2^2)\ln\left[\frac{s_4}{s_2}\right]+(s_3 s_4)^4(s_3^2-s_4^2)\ln\left[\frac{s_3}{s_4}\right],\nonumber
\end{eqnarray}
\begin{eqnarray}
  \varpi_2&=&(s_1 s_2)^2(s_3^4-s_4^4)\ln\left[\frac{s_1}{s_2}\right]+(s_1 s_3)^2(s_4^4-s_2^4)\ln\left[\frac{s_1}{s_3}\right]
  +(s_1 s_4)^2(s_4^2-s_3^4)\ln\left[\frac{s_1}{s_4}\right]\nonumber\\
  &+&(s_2 s_3)^2(s_1^4-s_4^4)\ln\left[\frac{s_2}{s_3}\right]+
  (s_2 s_4)^2(s_3^4-s_1^4)\ln\left[\frac{s_2}{s_4}\right]+(s_3 s_4)^2(s_1^4-s_2^4)\ln\left[\frac{s_3}{s_4}\right],\nonumber
\end{eqnarray}
\begin{eqnarray}
  \varpi_3&=&(s_1 s_2)^4(s_3^2-s_4^2)\ln\left[\frac{s_1}{s_2}\right]+(s_1 s_3)^4(s_4^2-s_2^2)\ln\left[\frac{s_1}{s_3}\right]
  +(s_1 s_4)^4(s_2^2-s_3^2)\ln\left[\frac{s_1}{s_4}\right]\nonumber\\
  &+&(s_2 s_3)^4(s_1^2-s_4^2)\ln\left[\frac{s_2}{s_3}\right]+
  (s_2 s_4)^4(s_3^2-s_1^2)\ln\left[\frac{s_2}{s_4}\right]+(s_3 s_4)^4(s_1^2-s_2^2)\ln\left[\frac{s_3}{s_4}\right],\nonumber
\end{eqnarray}
\begin{eqnarray}
  \lefteqn{\zeta_1=
  (s_2 s_3 s_4)^2(s_2^2-s_3^2)(s_2^2-s_4^2)(s_3^2-s_4^2)\ln\left[\frac{\gamma}{s_1}\right]}\nonumber\\
  &&{}+(s_1 s_3 s_4)^2(s_1^2-s_3^2)(s_1^2-s_4^2)(s_3^2-s_4^2)\ln\left[\frac{\gamma}{s_2}\right]\nonumber\\
  &&{}+(s_1 s_2 s_4)^2(s_1^2-s_2^2)(s_2^2-s_4^2)(s_1^2-s_4^2)\ln\left[\frac{\gamma}{s_3}\right]\nonumber\\
  &&{}+(s_1 s_2 s_3)^2(s_2^2-s_3^2)(s_1^2-s_2^2)(s_1^2-s_3^2)\ln\left[\frac{\gamma}{s_4}\right]\nonumber\\
  &&{}+(\gamma s_3 s_4)^2(\gamma^2-s_3^2)(\gamma^2-s_4^2)(s_3^2-s_4^2)\ln\left[\frac{s_1}{s_2}\right]\nonumber\\
  &&{}+(\gamma s_2 s_4)^2(\gamma^2-s_2^2)(\gamma^2-s_4^2)(s_2^2-s_4^2)\ln\left[\frac{s_1}{s_3}\right]\nonumber\\
  &&{}+(\gamma s_2 s_3)^2(\gamma^2-s_2^2)(\gamma^2-s_3^2)(s_2^2-s_3^2)\ln\left[\frac{s_1}{s_4}\right]\nonumber\\
  &&{}+(\gamma s_1 s_4)^2(\gamma^2-s_1^2)(\gamma^2-s_4^2)(s_1^2-s_4^2)\ln\left[\frac{s_2}{s_3}\right]\nonumber\\
  &&{}+(\gamma s_1 s_3)^2(\gamma^2-s_1^2)(\gamma^2-s_3^2)(s_1^2-s_3^2)\ln\left[\frac{s_2}{s_4}\right]\nonumber\\
  &&{}+(\gamma s_1 s_2)^2(\gamma^2-s_1^2)(\gamma^2-s_2^2)(s_1^2-s_2^2)\ln\left[\frac{s_3}{s_4}\right],\nonumber
\end{eqnarray}
\begin{eqnarray}
  \zeta_2&=&(\gamma s_1)^2[(s_2 s_3)^2+(s_2 s_4)^2+(s_3 s_4)^2](s_2^2-s_3^2)(s_2^2-s_4^2)(s_3^2-s_4^2)\ln\left[\frac{\gamma}{s_1}\right]\nonumber\\
  &+&(\gamma s_2)^2[(s_1 s_3)^2+(s_1 s_4)^2+(s_3 s_4)^2](s_1^2-s_3^2)(s_1^2-s_4^2)(s_3^2-s_4^2)\ln\left[\frac{\gamma}{s_2}\right]\nonumber\\
  &+&(\gamma s_3)^2[(s_1 s_2)^2+(s_1 s_4)^2+(s_2 s_4)^2](s_1^2-s_2^2)(s_2^2-s_4^2)(s_1^2-s_4^2)\ln\left[\frac{\gamma}{s_3}\right]\nonumber\\
  &+&(\gamma s_4)^2[(s_1 s_2)^2+(s_1 s_3)^2+(s_2 s_3)^2](s_2^2-s_3^2)(s_1^2-s_2^2)(s_1^2-s_3^2)\ln\left[\frac{\gamma}{s_4}\right]\nonumber\\
  &+&(s_1 s_2)^2[(\gamma s_3)^2+(\gamma s_4)^2+(s_3 s_4)^2](\gamma^2-s_3^2)(\gamma^2-s_4^2)(s_3^2-s_4^2)\ln\left[\frac{s_1}{s_2}\right]\nonumber\\
  &+&(s_1 s_3)^2[(\gamma s_2)^2+(\gamma s_4)^2+(s_2 s_4)^2](\gamma^2-s_2^2)(\gamma^2-s_4^2)(s_2^2-s_4^2)\ln\left[\frac{s_1}{s_3}\right]\nonumber\\
  &+&(s_1 s_4)^2[(\gamma s_2)^2+(\gamma s_3)^2+(s_2 s_3)^2](\gamma^2-s_2^2)(\gamma^2-s_3^2)(s_2^2-s_3^2)\ln\left[\frac{s_1}{s_4}\right]\nonumber\\
  &+&(s_2 s_3)^2[(\gamma s_1)^2+(\gamma s_4)^2+(s_1 s_4)^2](\gamma^2-s_1^2)(\gamma^2-s_4^2)(s_1^2-s_4^2)\ln\left[\frac{s_2}{s_3}\right]\nonumber\\
  &+&(s_2 s_4)^2[(\gamma s_1)^2+( s_3)^2+(s_1 s_3)^2](\gamma^2-s_1^2)(\gamma^2-s_3^2)(s_1^2-s_3^2)\ln\left[\frac{s_2}{s_4}\right]\nonumber\\
  &+&(s_3 s_4)^2[(\gamma s_1)^2+(\gamma s_2)^2+(s_1 s_2)^2](\gamma^2-s_1^2)(\gamma^2-s_2^2)(s_1^2-s_2^2)\ln\left[\frac{s_3}{s_4}\right],\nonumber
\end{eqnarray}

\begin{eqnarray}
  \zeta_3&=&(\gamma s_1)^4(s_2^2+s_3^2+s_4^2)(s_2^2-s_3^2)(s_2^2-s_4^2)(s_3^2-s_4^2)\ln\left[\frac{\gamma}{s_1}\right]\nonumber\\
  &+&(\gamma s_2)^4(s_1^2+s_3^2+s_4^2)(s_1^2-s_3^2)(s_1^2-s_4^2)(s_3^2-s_4^2)\ln\left[\frac{\gamma}{s_2}\right]\nonumber\\
  &+&(\gamma s_3)^4(s_1^2+s_2^2+s_4^2)(s_1^2-s_2^2)(s_2^2-s_4^2)(s_1^2-s_4^2)\ln\left[\frac{\gamma}{s_3}\right]\nonumber\\
  &+&(\gamma s_4)^4(s_1^2+s_2^2+s_3^2)(s_2^2-s_3^2)(s_1^2-s_2^2)(s_1^2-s_3^2)\ln\left[\frac{\gamma}{s_4}\right]\nonumber\\
  &+&(s_1 s_2)^4(\gamma^2+s_3^2+s_4^2)(\gamma^2-s_3^2)(\gamma^2-s_4^2)(s_3^2-s_4^2)\ln\left[\frac{s_1}{s_2}\right]\nonumber\\
  &+&(s_1 s_3)^4(\gamma^2+s_2^2+s_4^2)(\gamma^2-s_2^2)(\gamma^2-s_4^2)(s_2^2-s_4^2)\ln\left[\frac{s_1}{s_3}\right]\nonumber\\
  &+&(s_1 s_4)^4(\gamma^2+s_2^2+s_3^2)(\gamma^2-s_2^2)(\gamma^2-s_3^2)(s_2^2-s_3^2)\ln\left[\frac{s_1}{s_4}\right]\nonumber\\
  &+&(s_2 s_3)^4(\gamma^2+s_1^2+s_4^2)(\gamma^2-s_1^2)(\gamma^2-s_4^2)(s_1^2-s_4^2)\ln\left[\frac{s_2}{s_3}\right]\nonumber\\
  &+&(s_2 s_4)^4(\gamma^2+s_1^2+s_3^2)(\gamma^2-s_1^2)(\gamma^2-s_3^2)(s_1^2-s_3^2)\ln\left[\frac{s_2}{s_4}\right]\nonumber\\
  &+&(s_3 s_4)^4(\gamma^2+s_1^2+s_2^2)(\gamma^2-s_1^2)(\gamma^2-s_2^2)(s_1^2-s_2^2)\ln\left[\frac{s_3}{s_4}\right],\nonumber
\end{eqnarray}
\begin{eqnarray}
  \lefteqn{\zeta_4=(\gamma
  s_1)^6(s_2^2-s_3^2)(s_2^2-s_4^2)(s_3^2-s_4^2)\ln\left[\frac{\gamma}{s_1}\right]}\nonumber\\
  &&{}+(\gamma s_2)^6(s_1^2-s_3^2)(s_1^2-s_4^2)(s_3^2-s_4^2)\ln\left[\frac{\gamma}{s_2}\right]\nonumber\\
  &&{}+(\gamma s_3)^6(s_1^2-s_2^2)(s_2^2-s_4^2)(s_1^2-s_4^2)\ln\left[\frac{\gamma}{s_3}\right]\nonumber\\
  &&{}+(\gamma s_4)^6(s_2^2-s_3^2)(s_1^2-s_2^2)(s_1^2-s_3^2)\ln\left[\frac{\gamma}{s_4}\right]\nonumber\\
  &&{}+(s_1 s_2)^6(\gamma^2-s_3^2)(\gamma^2-s_4^2)(s_3^2-s_4^2)\ln\left[\frac{s_1}{s_2}\right]\nonumber\\
  &&{}+(s_1 s_3)^6(\gamma^2-s_2^2)(\gamma^2-s_4^2)(s_2^2-s_4^2)\ln\left[\frac{s_1}{s_3}\right]\nonumber\\
  &&{}+(s_1 s_4)^6(\gamma^2-s_2^2)(\gamma^2-s_3^2)(s_2^2-s_3^2)\ln\left[\frac{s_1}{s_4}\right]\nonumber\\
  &&{}+(s_2 s_3)^6(\gamma^2-s_1^2)(\gamma^2-s_4^2)(s_1^2-s_4^2)\ln\left[\frac{s_2}{s_3}\right]\nonumber\\
  &&{}+(s_2 s_4)^6(\gamma^2-s_1^2)(\gamma^2-s_3^2)(s_1^2-s_3^2)\ln\left[\frac{s_2}{s_4}\right]\nonumber\\
  &&{}+(s_3 s_4)^6(\gamma^2-s_1^2)(\gamma^2-s_2^2)(s_1^2-s_2^2)\ln\left[\frac{s_3}{s_4}\right].\nonumber
\end{eqnarray}



\begin{thebibliography}{99}
\bibitem{knigaMenskogo} M.B.~Mensky,
{\it Quantum Mesurements and Decogerence} (Kluwer Academic Publishers, 2000).

\bibitem{Fried} H.~Friedrich and D.~Wintgen, Phys. Rep. {\bf 183}, 37 (1989);
A.~Holle, J.~Main, G.~Wiebusch, H.~Rottke and
K.H.~Welge, Phys. Rev. Lett. {\bf 61}, 161 (1988).

\bibitem{Chuvil} S.D.~Kurgalin, I.S.~Okunev, T.V.~Chuvilskaya, and
Yu.M.~Tchuvil'sky, Yadernay Fizika {\bf 68}, 2042 (2005).

\bibitem{PS} L.~Jacak, P.~Hawrylak, and A.~Wojs,
{\it Quantum Dots} (Springer-Verlag, Berlin, Heidelberg, New York, 1997).

\bibitem{Naz}W.D.~Heiss and R.G.~Nazmitdinov, Phys.Rev. B {\bf 55}, 16310 (1997);
Pis'ma v ZhETF {\bf 68}, 870 (1998); M.~Dineykhan and R.G.~Nazmitdinov,
Phys. Rev. B {\bf 55}, 13707 (1997); R.G.~Nazmitdinov, N.S.~Simonovic, and J.M.~Rost,
Phys.Rev. B {\bf 65}, 155307 (2002).

\bibitem{Naz2}
Yu.~Demidenko, A.~Kuzyk, V.~Lozovski, and O.~Tretyak,
J. Phys. C {\bf 16}, 543 (2004); A.~Matulis and E.~Anisimovas,
J. Phys. C {\bf 17}, 3851 (2005).

\bibitem{Langer} L.~Langer {\it et al}., Phys. Rev. Lett. {\bf 76}, 479 (1996);
O.~Bourgeois {\it et al}., arXiv:cond-mat/9901045;
E.L.~Nagaev, Uspehi Fizicheskih Nauk {\bf 166}, 833 (1996).

\bibitem{Ginz} V.L.~Ginzburg and A.V.~Gurevich,
Uspehi Fizicheskih Nauk {\bf LXX}, 201 (1960).

\bibitem{Kampen1}  N.G. van Kampen,
{\it Stochastic Processes in Physics and Chemistry}
(North-Holland, Amsterdam, 1981).
\bibitem{Kampen2}  C.W. Gardiner,
{\it Quantum Noise} (Springer, Berlin, 1991).
\bibitem{Kampen3}  U. Weiss, {\it Quantum Dissipative Systems}  (World Scientific, Singapore, 1999).
\bibitem{Kampen4}   D. Zubarev, V. Morozov, and
G. R\"{o}pke, \textit{Statistical mechanics of nonequilibrium processes}, vol. 2 (Academie Verlag, Berlin, 1997).
\bibitem{Kampen5}   H.J. Carmichael, \textit{An open system approach to quantum optics}, Springer, Berlin (1993).
\bibitem{Kampen6}   Yu.L. Klimontovich, \textit{Statistical theory of open systems} (Kluwer Academic Publishers, Dordrecht, 1995).
\bibitem{LEG} A.O.~Caldeira and A.J.~Leggett, {Phys. Rev. Lett.} {\bf 46}, 211 (1981);
{Phys. Rev. Lett.} {\bf 48}, 1571 (1982); {Ann. Phys. (N.Y.)} {\bf 149}, 374 (1983).

\bibitem{Dodonov}  V.V.~Dodonov and O.V.~Man'ko,
Sov. J. Theoretical and Mathematical Physics {\bf 65}, 1 (1985);
\bibitem{DM} V. V. Dodonov and
V. I. Man'ko, {\it Density Matrices and Wigner Functions of Quasiclassical Quantum Systems}
(Proc. Lebedev Phys. Inst. of Sciences, Vol. {\bf  167}, A. A. Komar, ed.)
(Nova Science, Commack, N. Y., 1987).

\bibitem{Hu1} G.W. Ford, J.T. Lewis, R.F. O'Connell, Phys. Rev. A {\bf 36}, 1466 (1987); A {\bf 37}, 4419 (1988).

\bibitem{Hu2} G.Y. Hu and R.F. O'Connell, Physica A {\bf 151}, 33 (1988); Phys. Rev. B {\bf 36}, 5798 (1987).

\bibitem{Ma} Y. Marathe, Phys. Rev. A {\bf 39}, 5927 (1989).

\bibitem{Katia}   K. Lindenberg and B. J. West,
 {\it The Nonequilibrium Statistical Mechanics of Open and Closed Systems}
 (VCH Publishers, Inc., New York,  1990);
 K. Lindenberg and B. J. West,
Phys. Rev. A {\bf 30}, 568 (1984).

\bibitem{Isar} A. Isar, A. Sandulescu, H. Scutaru, E. Stefanescu, and W. Scheid,
Int. J. Mod. Phys. E {\bf 3}, 635 (1994).

\bibitem{Ford}  X.L.~Li, G.W.~Ford, R.F.~O'Connell,
Phys.~Rev. A {\bf 41}, 5287 (1990); ibid {\bf 42}, 4519 (1990);
Physica A {\bf 193}, 575 (1993).
\bibitem{Ford1}  X.L.~Li, G.W.~Ford, R.F.~O'Connell,
Phys.~Rev. E {\bf 53}, 3359 (1996).

\bibitem{In}  S.~Dattagupta and J.~Singh, Phys. Rev.
Lett. {\bf 79}, 961 (1997).




\bibitem{M110}
Th.M. Nieuwenhuizen and A.E. Allahverdyan,   Phys. Rev. E  {\bf 66}, 036102  (2002).

\bibitem{Kanokov}  Z. Kanokov, Yu.V. Palchikov, G.G. Adamian,
N.V. Antonenko, and W. Scheid,  Phys.~Rev. E {\bf 71},
016121 (2005).

\bibitem{Kanokov2}
Sh.A.~Kalandarov,
Z. Kanokov,  G.G. Adamian, and
N.V. Antonenko,   Phys.~Rev. E {\bf 75}, 0311115 (2007).

\bibitem{PA}
I.B.~Abdurakhmanov,
  G.G. Adamian,
N.V. Antonenko, and Z. Kanokov,  Physica A {\bf 508}, 613 (2018);
Eur. Phys. J.   B {\bf 91} (2018) in print.

\bibitem{Lac11} V.V. Sargsyan, Z. Kanokov, G.G. Adamian, and N.V. Antonenko,
Phys. Part. Nuclei {\bf 41}, 175 (2010).
\bibitem{Lac13}
K. Wen, F. Sakata, Z.-X. Li, X.-Z. Wu, Y.-X. Zhang, and S.-G. Zhou,
Phys. Rev. Lett. {\bf 111}, 012501 (2013).
\bibitem{Lac14} D. Lacroix, V.V. Sargsyan, G.G. Adamian, and N.V. Antonenko,  Eur. Phys. J. B {\bf 88}, 89 (2015).





\end{thebibliography}
\end{document}